\documentclass{article}
\usepackage{amssymb,latexsym,amsfonts,amsmath,amsthm}
\usepackage[dvips]{graphics}
\renewcommand{\d}{\delta}

\newcommand{\C}{\mathbb{C}}

\newcommand{\Z}{\mathbb{Z}}

\newcommand{\R}{\mathbb{R}}

\newcommand{\N}{\mathbb{N}}

\newcommand{\Hil}{\mathcal{H}}
\renewcommand{\H}{\mathcal{H}}
\newcommand{\B}{\mathfrak{b}}

\newcommand{\unity}{\boldsymbol{1}}

\newcommand{\ket}[1]{\left\vert #1\right\rangle}

\newcommand{\spec}{\operatorname{spec}}
\newcommand{\infspec}{\operatorname{infspec}}
\newcommand{\floor}[1]{\lfloor{#1}\rfloor}
\newcommand{\ip}[2]{\left(#1, #2\right)}

\theoremstyle{plain}
\newtheorem{thm}{THEOREM}[section]
\newtheorem{lemma}[thm]{LEMMA}
\newtheorem{cl}[thm]{COROLLARY}
\newtheorem{prop}[thm]{PROPOSITION}

\theoremstyle{definition}

\theoremstyle{remark}
\newtheorem{remark}[thm]{REMARK}
\newtheorem{remarks}[thm]{REMARKS}

\newcommand{\be}{\begin{equation}}
\newcommand{\ee}{\end{equation}}
\newcommand{\bea}{\begin{eqnarray}}
\newcommand{\eea}{\end{eqnarray}}
\newcommand{\beax}{\begin{eqnarray*}}
\newcommand{\eeax}{\end{eqnarray*}}

\newcommand{\HW}{\textrm{hw}}

\renewcommand{\S}{\mathbb{S}}
\newcommand{\per}{{\rm cyc}}

\begin{document}

\title{Droplet Excitations for the Spin-$1/2$ 
XXZ Chain \\[5pt]
with Kink Boundary Conditions}

\author{B.~Nachtergaele$^{1}$, W.~Spitzer$^{2}$ and S.~Starr$^{3}$\\[10pt]
{$^{1}$Department of Mathematics}\\
{University of California, Davis}\\
{One Shields Avenue}\\
{Davis, CA 95616-8366, USA}\\
\texttt{bxn@math.ucdavis.edu}\\[5pt]
 {$^{2}$Department of Mathematics}\\
{University of British Columbia}\\
{Room 121, 1984 Mathematics Road}\\
{Vancouver, B.C., Canada V6T 1Z2}\\
\texttt{spitzer@math.ubc.ca}\\[5pt]
{$^{3}$ UCLA Mathematics Department}\\
{Box 951555}\\
{Los Angeles, CA 90095-1555, USA}\\
\texttt{sstarr@math.ucla.edu}
}

\date{August 24, 2005}
\maketitle

\begin{abstract} 
We give a precise definition for excitations consisting of a droplet of size
$n$ in the XXZ chain with various choices of boundary conditions, including
kink boundary conditions and prove that, for each $n$, the droplet energies
converge to a boundary condition independent value in the thermodynamic limit.
We rigorously compute an explicit formula for this limiting value using the
Bethe Ansatz.
\end{abstract}

\renewcommand{\thefootnote}{$1$}
\footnotetext{Work partially supported by U.S. National Science
Foundation grant under Grant \# DMS-0303316.}
\renewcommand{\thefootnote}{$2$}
\footnotetext{Work partially supported by the Natural Sciences and
    Engineering Research Council of Canada}

\thanks{Copyright \copyright\ 2005 by the authors. This article may be
reproduced in its entirety for non-commercial purposes.}

\tableofcontents

\section{Introduction}

In this paper we study the low-energy spectrum of the one-dimensional
spin-$1/2$ ferromagnetic XXZ Heisenberg Hamiltonian in the thermodynamic limit.
The specific questions we are interested in concern the excitations that
describe droplets, i.e., finite domains of reversed magnetization. The simplest
case is where the infinite chain is in one of its two translation-invariant
ground states with all spins parallel or antiparallel to the $z-$axis. A
droplet excitation is then a state with $n$, $n\geq 1$, opposite spins that
form, up to quantum fluctuations, a compact cluster which moves through the
system as a unit. Since the model also has kink and antikink ground states
\cite{Matsui1996, Koma1998} in which two halfs of the chain have opposite
magnetization with a transition region in between, it is interesting to ask
about droplet excitations with respect to such a ground state. This raises some
interesting questions about how to define a droplet excitation in this case and
how to approximate them by excited states in finite volume obtained by imposing
boundary conditions or a constraint such as a particular value of the total
magnetization.

We will consider the spin-$1/2$ XXZ chain of length $L$ with Hamiltonian
$$   
H_{[1,L]}\, =\, - \sum_{x=1}^{L-1}  \frac{1}{\Delta} \left[S_x^1\, 
S_{x+1}^1 + S_x^2\, S_{x+1}^2\right] 
+ S_x^{3}\, S_{x+1}^{3} ,
$$
and study $H_{[1,L]} + h_{1,L}$, where $h_{1,L}$ is one of the following 
three choices of boundary term: 
\begin{align}
&\mbox{periodic b.c.:\ }  &h_{1,L} = - \frac{1}{\Delta} \left[S_1^1\, S_L^1 + S_1^2\, S_L^2\right] 
- S_1^{3}\, S_L^{3}\\
&\mbox{droplet b.c.:\ } &h_{1,L} = -\delta\left(S^3_1 + S^3_L\right)\\
&\mbox{kink b.c.:\ } &h_{1,L} = -\alpha\left(S^3_1 - S^3_L\right)
\end{align}

See Section \ref{sec:main_results} for suitable choices of the constants
$\alpha$ and $\delta$  as well as other definitions. In the first two cases one
can define the droplet energy by  restricting the Hamiltonian to invariant
subspaces of fixed total third component of the spin. Let $\lambda(n)$ denote
the smallest eigenvalue of the Hamiltonian under consideration restricted to
the subspace of states with $n$ down spins and $L-n$ up spins, $0\leq n\leq L$,
which is called the space of $n-$magnon states. For the case of periodic or
droplet boundary conditions, the energy  of a droplet of size $n$ is then
defined  to be $\lambda(n)-\lambda(0)$. For the model with kink boundary
conditions this strategy does not work, since $\lambda(n)$ is attained in a
kink ground state and is independent of $n$. The kink ground states form a
multiplet of maximal $SU_q(2)$ spin: $S_{\rm max}=L/2$. It turns  out that the
correct subspace to define droplet excitations is the subspace of fixed total
spin $S=S_{\rm max}-n$. The mathematical explanation for this definition lies
in the existence of a linear isomorphism between the space of $n-$magnon states
and the ``highest weight'' vectors of ``weight'' $S_{\rm max}-n$. The quotation
marks are necessary here, since the isomorphism only exists for the infinite
chain and the weights are not well-defined (infinite). See Section \ref{sec:R}
for the definition of this isomorphism, which we will denote by $R$. $R$ is a 
bounded invertible operator that intertwines the Hamiltonians with kink and
droplet boundary conditions on the infinite chain. Another, more physical
interpretation, is that the total spin quantum number associated with a droplet
of size $n$ is $n/2$, in agreement with the case $n=1$, more commonly known as
spin waves \cite{Faddeev1981}

Our main results can be summarized in words as follows: the droplet energies
defined with the different boundary conditions above all converge to the same
value in thermodynamic limit and that value can be computed exactly by the
Bethe Ansatz. The result is given in Theorem \ref{thm:main}. In the proof of
this theorem we use Perron-Frobenius type arguments to turn the Bethe Ansatz
calculation into rigorous mathematics. For the precise definitions and
mathematical statements we refer the reader to Section  \ref{sec:main_results}.

The main motivation for this study is to complete our understanding of the
low-lying spectrum of the XXZ chain, which is important for a variety of
problems involving the dynamics. As a by-product we have also come a step
closer to a complete proof of the completeness of the Bethe Ansatz in the 
thermodynamic limit.

\section{Set-up and main results}\label{sec:main_results}

\subsection{The kink Hamiltonian}

For $L \in \N_+$, consider a spin chain on the sites of $[1,L]
\subset \Z$. The Hilbert space is $\Hil = \Hil_{[1,L]} =
\bigotimes_{x \in [1,L]} \Hil_x$, where $\Hil_x$ is a
two-dimensional Hilbert space for each $x \in [1,L]$. We take an
orthonormal basis of $\Hil_x$ to be the Ising basis
$\{\ket{\uparrow},\ket{\downarrow}\}$. The spin-$1/2$
representation of $\textrm{SU}(2)$ is defined on $\C^2$ through
the matrices
\begin{equation*}
    S^{1} = \begin{bmatrix} 0 & 1/2 \\ 1/2 & 0 \end{bmatrix}\,
    ,\qquad
    S^{2} = \begin{bmatrix} 0 & -i/2 \\ i/2 & 0 \end{bmatrix}\,
    ,\qquad
    S^{3} = \begin{bmatrix} 1/2& 0 \\ 0 & -1/2 \end{bmatrix}\, ,
\end{equation*}
in the $\{\ket{\uparrow},\ket{\downarrow}\}$ basis. For each $x
\in [1,L]$ and $i\in\{1,2,3\}$ we have the operators $S_x^i$ on
$\Hil$ where $S^i$ acts on $\Hil_x$ and is tensored with $\unity$
on $\Hil_y$ for all $y\neq x$.

The XXZ model is the Hamiltonian
\begin{align}
    H\, &=\, \sum_{x=1}^{L-1} h_{x,x+1}\\
    \label{eq:bdryless}
  h_{x,x+1}\, &=\, \frac{1}{4} \unity - S_x^{3}\, S_{x+1}^{3} -
  \frac{1}{\Delta} \left[S_x^1\, S_{x+1}^1 + S_x^2\, S_{x+1}^2\right]\,
  ,
\end{align}
with $\Delta\geq 1$. Since we consider $1/\Delta$, it is allowable
that $\Delta=+\infty$. Let $q$ be the number in $[0,1]$ such that
$\Delta = (q+q^{-1})/2$. Then a modification of this
Hamiltonian is the so-called kink Hamiltonian
\begin{equation}
    H^{\rm k}\, =\, H - \frac{\alpha}{2} S_1^3 + \frac{\alpha}{2} S_L^3\, ,
\end{equation}
where $\alpha$ is the constant
\begin{equation}
    \alpha\, =\, \frac{1-q^2}{1+q^2}\, .
\end{equation}
When $q=1$, this gives the isotropic Heisenberg model without
boundary fields. It is useful to incorporate the alternating
boundary fields into the nearest-neighbor interactions,
\begin{equation}
    h^{\rm k}_{x,x+1}\, =\,
    \frac{1}{4} \unity - S_x^{3} S_{x+1}^{3} - \frac{1}{\Delta}
    \left(S_x^{1} S_{x+1}^{1} + S_{x}^{2}
    S_{x+1}^{2}\right)
    - \frac{\alpha}{2} \left(S_{x}^{3} - S_{x+1}^{3}\right)\, .
\end{equation}
Then the kink Hamiltonian can be written as
\begin{equation}
    H^{\rm k}\, =\, \sum_{x=1}^{L-1} h_{x,x+1}^{\rm k}\, .
\end{equation}
When we want to emphasize the chain for the Hamiltonian, we will
write $H_{[1,L]}$ for $H$ and $H^{\rm k}_{[1,L]}$ for $H^{\rm k}$.

There are three important operators commuting with each
nearest-neighbor kink interaction, separately. The first is
\begin{equation}
    S^3_{[1,L]}\, =\, \sum_{x=1}^{L} S_x^3\, .
\end{equation}
This is the usual total-magnetization operator for representations
of $\textrm{SU}(2)$. The other two operators are $q$-versions of
the total raising and lowering operators
\begin{align}
  S_{[1,L]}^+\, &:=\, \sum_{x=1}^{L} q^{-2(S_1^3 + \dots + S_{x-1}^3)}\, 
  S_x^+\, ,\\
  S_{[1,L]}^-\, &:=\, \sum_{x=1}^{L} S_x^-\, q^{2(S_{x+1}^3 
  + \dots + S_L^3)}\,
  .
\end{align}
Note that these operators are only well-defined when $0<q\leq 1$.
These three operators together give a representation of the
quantum group $\textrm{SU}_q(2)$. (For readers unfamiliar with quantum groups, 
we will present all the details necessary for our results.)

The total magnetization eigenvalues are $\{L/2-n\, :\, n=0,1,\dots,L\}$. The
eigenspace for the eigenvalue $L/2-n$ will be denoted as $\Hil(n)$. It is an
invariant subspace for $H^{\rm k}$. For $n< L/2$, $S^-_{[1,L]}$ maps $\Hil(n)$
isomorphically onto its image in $\Hil(n+1)$. (For a proof of this and other
facts about the representations of $\textrm{SU}_q(2)$ c.f.\ \cite{Kassel1995}.)
Moreover, this image is an invariant subspace of $H^{\rm k}$. Therefore, so is
its orthogonal complement. For $1\leq n\leq \floor{L/2}$, we may define
$\Hil^{\HW}(n)$ as the subspace of $\Hil(n)$ such that
\begin{equation*}
    \Hil(n)\, =\, \Hil^{\HW}(n) \oplus S^{-}_{[1,L]} \Hil(n-1)\,.
\end{equation*}
Define $\Hil^{\HW}(0)=\Hil(0)$. Then, $\Hil^{\HW}(n)$ consists of
vectors in $\Hil$ which have total $\textrm{SU}_q(2)$ spin equal
to $L/2-n$, and which are highest-weight vectors in the sense that
$S_{[1,L]}^+$ annihilates each such vector. (C.f., 
\cite{Kassel1995} for more information.) As noted above,
$\Hil^{\HW}(n)$ is an invariant subspace for each $n
=0,\dots,\floor{L/2}$. 

One can define a subspace $\Hil^{\textrm{sd}}(n)$ to be the set of all vectors
whose total $\textrm{SU}_q(2)$ spin is $L/2-n$. We would call this the
``$n$-spin deviate'' subspace because the total spin deviates from the maximum
possible value of $L/2$ by $n$. (Total spin is a function of the Casimir
operator, which generates the center of the algebra of $\textrm{SU}_q(2)$,
which matches the usual notion for $\textrm{SU}(2)$ total spin when $q=1$.)
Since the total spin operator commutes with $S_{[1,L]}^3$, one can define
subspaces $\Hil^{\textrm{sd}}(n,k)$ which are subspaces of
$\Hil^{\textrm{sd}}(n)$ with $S^3_{[1,L]}$ eigenvalue equal to $L/2-k$. These
subspaces are trivial unless $n\leq k\leq L-n$. Therefore 
\begin{equation*}
\Hil^{\textrm{sd}}(n)\, =\, \bigoplus_{k=0}^{L-2n} 
\Hil^{\textrm{sd}}(n,n+k)\, .
\end{equation*}
Also, $\Hil^{\textrm{sd}}(n,n) = \Hil^{\HW}(n)$. For $k=1,\dots,L-2n$ one has 
$\Hil^{\textrm{sd}}(n,n+k)=(S^-_{[1,L]})^k \Hil^{\HW}(n)$, and this is an 
isomorphic image. Since $S^-_{[1,L]}$ commutes with $H^{\rm k}$, each subspace 
is an invariant subspace for $H^{\rm k}$. 

It is natural to define
\begin{equation*}
E(L,n)\, =\, \infspec\big(H^{\rm k} \restriction \Hil^{\textrm{sd}}(n)\big)\, ,
\end{equation*}
which is the minimum energy of $H^{\rm k}$ ranging over all vectors in the $n$-spin deviate subspace. We make explicit reference to the length of the chain $[1,L]$ in this notation. On the other hand, $H^{\rm k}$ commutes with $S^-_{[1,L]}$ and one can generate all of $\Hil^{\textrm{sd}}(n)$ by acting on $\Hil^{\HW}(n)$ by $S^-_{[1,L]}$ some number of times.
Therefore, it is clear that
\begin{equation*}
\infspec\big(H^{\rm k} \restriction \Hil^{\textrm{sd}}(n)\big)\,
=\, \infspec\big(H^{\rm k} \restriction \Hil^{\HW}(n)\big)\, .
\end{equation*}
This is the definition we will use henceforth. We now define $H^{\rm k}_n = 
H^{\rm k}\restriction\Hil^{\HW}(n)$. So, $E(L,n) = \infspec(H^{\rm k}_n)$. 
By Theorem 1.4 in \cite{Nachtergaele2004},
\begin{equation}
    E(L,0)\, \leq\, E(L,1)\, \leq\, \dots\,  \leq\, E(L,\floor{L/2})\, ,
\end{equation}
for each finite $L$. By Proposition 4.1 in that same paper we know
that all the inequalities are strict, at least as long as $0<q\leq
1$. For $q=0$ one cannot define $E(L,n)$ because there is no
quantum group representation, but taking the limit as $q\to 0^+$
gives $E(L,1)=\dots=E(L,\floor{L/2})=1$ which satisfy the
inequalities, but not strictly.

Moreover, by Proposition 7.1 in that paper, the sequence
$(E(L,n)\, :\, L \geq 2n)$ is decreasing in $L$. Therefore, the
limit $\lim_{L\to\infty} E(L,n)$ necessarily exists. Obviously one has
inequalities 
\begin{equation*}
\lim_{L\to\infty} E(L,n+1)\, \geq\, \lim_{L\to\infty} E(L,n)\, ,
\end{equation*}
which are derived from the fact that $E(L,n+1)>E(L,n)$ for every finite $L$. 
But note that one cannot automatically conclude that the inequality is strict 
in the limit.
Whether or not this is so is a natural question.
One might hope to resolve this question by finding an explicit formula for 
the limits.
This is the first main result of the paper.
\begin{thm}
\label{thm:main} For all $n \in \N$, and $0<q<1$
\begin{equation}
\lim_{L\to\infty} E(L,n)\, =\, \frac{(1-q^2)(1-q^n)}{(1+q^2)(1+q^n)}\, .
\label{enq}\end{equation}
\end{thm}

For two values of $n$ a formula for $\lim_{L\to\infty} E(L,n)$ was previously 
known. For $n=0$, one obtains the ground state energy. It  is well-known that 
the ground state energy is $E(L,0)=0$ for all finite $L$, in fact the 
Hamiltonian was constructed to satisfy this condition. For $n=1$, $E(L,n)$ 
measures the spectral gap. The formula for this was calculated for all finite 
$L$ in \cite{Koma1997}. The value of the limit is  $\lim_{L\to\infty} 
E(L,1) = 1 - \Delta^{-1}$. Since $\Delta=\frac{1}{2} (q+q^{-1})$, this is 
easily seen to agree with the result of the theorem.

\begin{remark} A different method for bounding $E(1)=\lim_{L\to\infty} E(L,1)$ 
was given in
\cite{Nachtergaele1996}. That method is based on the 
martingale method, which proved useful in interacting particle systems 
\cite{Lu1993}. Some inequalities of \cite{Nachtergaele1996} were made sharper 
in \cite{Spitzer2003}. In particular, this led to an independent derivation 
of $E(1)$. \end{remark}

\begin{remark}
\label{rem:cont}
Let $E_q(L,n)$ and $E_q(n)=\lim_{L\to\infty} E_q(L,n)$ be the relevant 
$q$-dependent quantities for $q \in [0,1]$. Using properties of the 
functions $E_q(L,n)$ (in particular monotonicity in $q$, c.f., Remark 
\ref{rem:mono}) one finds
\begin{equation*}
\lim_{r\to q^+} \lim_{L\to  \infty} E_r(L,n)\, 
=\, \lim_{L \to \infty} E_{q}(L,n)\, 
\leq\, \lim_{r\to q^-} \lim_{L\to \infty} E_r(L,n)\, ,
\end{equation*} 
whenever the relevant limits exist.  This can be used to recover the 
(obvious) fact that $E_q(n)=1$ for $q=0$ and $n\geq 1$. Also it can be 
used to obtain the upper bound $E_q(n)\leq 0$ for $q=1$, which matches 
the obvious lower bound. For $q=1$ direct spin-wave trial functions can 
also be used to verify $\lim_{L\to\infty} E_q(L,n)=0$ directly. (In fact, 
in \cite{Koma1997} the finite-size scaling was calculated for $E_q(L,n)$ 
when $q=1$ and $n=1$.)
In our proof, we use 
the fact that $0<q<1$. Therefore, we will make this assumption, henceforth.
\end{remark}

\subsection{Droplet Hamiltonians}

In \cite{Nachtergaele2001}, two of the authors investigated
low-energy vectors for three different droplet-type Hamiltonians
based on the XXZ model. For $\delta \in \R$ one may define
\begin{equation}\label{def:droplet}
    H^{\delta}_{[1,L]}\, =\,  H_{[1,L]}  + \frac{\delta}{2}\, (\unity - S_1^3 -
    S_L^3)\, ,
\end{equation}
on $\Hil_{[1,L]}$. We call this the ``droplet'' Hamiltonian. Also on
$\Hil_{[1,L]}$ the spin chain with periodic
boundary conditions (spin ring) is defined as
\begin{equation}\label{def:cyc}
    H^{\per}_{[1,L]}\, =\, H_{[1,L]} + h_{1,L}\, .
\end{equation}
We call this the ``cyclic'' Hamiltonian.
Neither of these Hamiltonians has the full $\textrm{SU}_q(2)$
symmetry. But they both have the symmetry of $S^3_{[1,L]}$. We
define
\begin{align}
    E^\delta(L,n)\, &:=\, \infspec\big(H^{\delta}_{[1,L]} \restriction
    \Hil_{[1,L]}(n)\big)\quad \textrm{and}\label{def:Edroplet}\\
    E^{\per}(L,n)\, &:=\, \infspec\big(H^{\per}_{[1,L]} \restriction
    \Hil_{[1,L]}(n)\big)\label{def:Ecyc}\, ,
\end{align}
for each $L\geq n$. 

Finally, let us recall that
one of the ground-states of the infinite-volume
Hamiltonian is the all-up-spin state. The  GNS representation for this state 
will be constructed 
in Section \ref{Section 3}. For now, let us write
$(\Hil_{\Z},\omega_{\Z},H_{\Z})$ for the GNS representation. There
are subspaces $\Hil_{\Z}(n)$ which are invariant for the
Hamiltonian, and such that $\Hil_{\Z} = \bigoplus_{n=0}^{\infty}
\Hil_{\Z}(n)$. These are the $n$-magnon subspaces. We define
\begin{equation}
    E_{\Z}(n)\, :=\, \infspec\big(H_{\Z} \restriction
    \Hil_{\Z}(n)\big)\, ,
\end{equation}
for each $n$. 

The second main result of the paper is the
following.
\begin{thm}
For $\delta\geq 1$,
\begin{equation}
    \lim_{L\to\infty} E^{\delta}(L,n)\, =\, \lim_{L \to \infty} 
    E^{\per}(L,n)\, =\, E_{\Z}(n)\, =\, \lim_{L\to\infty} E(L,n)\, .
\end{equation}
\end{thm}

\section{Droplet energies in the infinite chain}
\label{Section 3}

In this section we will give a precise definition for $E_{\Z}(n)$,
and calculate it for all $n \in \N$. 

\subsection{Set-up}

We will start by constructing the GNS Hilbert space for the all-up-spin ground
state of the infinite XXZ chain. Instead of following the usual GNS
construction, for this special case one can define the representation directly.
We define the GNS Hilbert space, $\Hil_{\Z}$ as the direct sum of Hilbert
spaces $\Hil_{\Z}(n)$ for $n\in \N = \{0,1,2,\dots\}$. Each of these subspaces
are $\ell^2$-spaces on countable sets. Let $\mathcal{X}_0=\{\emptyset\}$.
Define $\Hil_{\Z}(0) = \ell^2(\mathcal{X}_0)$, which is a 1-dimensional space.
For $n \in \N_+$, let $\mathcal{X}_n=\{\boldsymbol{x} \in \Z^n\, :\,
x_1<\dots<x_n\}$, and define $\Hil_{\Z}(n)=\ell^2(\mathcal{X}_n)$. This defines
$\Hil_{\Z}$. Given $n \in \N$ and $\boldsymbol{x} \in \mathcal{X}_n$, define
$\delta_{\boldsymbol{x}} \in \ell^2(\mathcal{X}_n)$ so that
$\delta_{\boldsymbol{x}}(\boldsymbol{y})
=\delta_{\boldsymbol{x},\boldsymbol{y}}$.
These define the natural orthonormal basis. Note that the basis for
$\Hil_{\Z}(0)$ is denoted $\delta_{\emptyset}$. Physically, the $x_i$ are just
the positions of the down spins.

We should next define operators on $\Hil_{\Z}$ satisfying the same commutation 
relations as the spin-matrices $S_x^i$ for $i=1,2,3$ from the last section, 
except now for all $x\in \Z$, not just a finite set $x\in\{1,\dots,L\}$.
(These generate the $C^*$-algebra on which the infinite XXZ Hamiltonian 
operates by the Heisenberg dynamics.)
For each $x \in \Z$, there is a representation of $\textrm{SU}(2)$
on $\Hil_{\Z}$, given as follows.
For each $n \in \N$, if $\boldsymbol{x} \in \mathcal{X}_n$, then:
\begin{itemize}
\item
$S_x^3 \delta_{\boldsymbol{x}} = m(x,\boldsymbol{x}) \delta_{\boldsymbol{x}}$, 
where $m(x,\boldsymbol{x})$ equals $+\frac{1}{2}$ if $x \in \{x_1,\dots,x_n\}$ 
and $-\frac{1}{2}$ otherwise;
\item
$S_x^- \delta_{\boldsymbol{x}}$ equals $0$ if $x \in \{x_1,\dots,x_n\}$, and 
otherwise it equals $\delta_{\boldsymbol{y}(x,\boldsymbol{x})}$ where 
$\boldsymbol{y}(x,\boldsymbol{x}) \in \mathcal{X}_{n+1}$ is 
\begin{equation*}
\boldsymbol{y}(x,\boldsymbol{x}) \, =\, (x_1,\dots,x_k,x,x_{k+1},\dots,x_n)
\, ,
\end{equation*}
    for that $k\in\{1,\dots,n\}$ such that $x_k<x<x_{k+1}$ 
    (considering $x_0=-\infty$ and $x_{n+1}=+\infty$);
\item
$S_x^+ \delta_{\boldsymbol{x}}$ equals $0$ unless $x \in \{x_1,\dots,x_n\}$, 
and in that case it equals $\delta_{\boldsymbol{z}(x,\boldsymbol{x})}$ where 
$\boldsymbol{z}(x,\boldsymbol{x}) \in \mathcal{X}_{n-1}$ is
\begin{equation*}
\boldsymbol{z}(x,\boldsymbol{x}) \, =\, (x_1,\dots,x_{k-1},x_{k+1},\dots,x_n)
\, ,
\end{equation*}
for that $k\in\{1,\dots,n\}$ such that $x=x_k$.
\end{itemize}
This is similar to the Fock space representation of the CCR algebra, except 
that there is a restriction to have at most one particle per site. Therefore, 
this is sometimes called the hard-core Bose gas. (It is also related to a 
Fock space representation of the CAR algebra using the Jordan-Wigner 
transformation.)

The cyclic GNS vector is the vacuum vector, $\delta_{\emptyset} 
\in \Hil_{Z}(0)$.
Then we may define the GNS Hamiltonian as 
\begin{equation}
  H_{\Z}\, =\, \sum_{x \in \Z} h_{x,x+1}\, ,
\end{equation}
where the interactions have the same formula as in
(\ref{eq:bdryless}), but relative to the present representation.
As in the finite case, each $n$-magnon subspace is an invariant
subspace for the Hamiltonian. It will be convenient to adopt a
notation for the restriction to the $n$-magnon subspace
\begin{equation}
    H_{\Z}(n)\, :=\, H_{\Z} \restriction \Hil_{\Z}(n)\,.
\end{equation}
We define the droplet energies
\begin{equation}
    E_{\Z}(n)\, =\,
    \infspec\, H_{\Z}(n)\, ,
\end{equation}
for each $n \in \N$. The main purpose of this section is to prove
the following result.

\begin{prop}
\label{prop:infspec} For each $n \in \N$,
\begin{equation}
\label{eq:form}
    E_{\Z}(n)\, =\, \frac{(1-q^2)(1-q^n)}{(1+q^2)(1+q^n)}\, .
\end{equation}
\end{prop}

The result is trivial for $n=0$. Henceforth, we will consider
$n>0$. Let $\boldsymbol{e}_1,\dots,\boldsymbol{e}_n$ be the
coordinate unit vectors in $\Z^n$. Then $H_{\Z}(n)$ can be represented as a 
discrete integral operator by a kernel
\begin{equation*}
H_{\Z}(n) f(\boldsymbol{x})\, =\, \sum_{\boldsymbol{y} \in \mathcal{X}_n} 
K_n(\boldsymbol{x},\boldsymbol{y})\, f(\boldsymbol{y})\, ,
\end{equation*}
for all $\boldsymbol{x} \in \mathcal{X}_n$.
The kernel is given by
\begin{equation}
    K_n(\boldsymbol{x},\boldsymbol{y})\,
    =\, \sum_{k=0}^n K_n^{(k)}(\boldsymbol{x},\boldsymbol{y})\, ,
\end{equation}
where:
\begin{equation*}
    K_n^{(0)}(\boldsymbol{x},\boldsymbol{y})\,
    =\, \frac{1}{2}\, \delta_{\boldsymbol{y},\boldsymbol{x}}
    - \frac{1}{2\Delta}\, \delta_{\boldsymbol{y},\boldsymbol{x}-
    \boldsymbol{e}_1}\, ;
\end{equation*}
for $k=1,\dots,n-1$,
\begin{equation*}
    K_n^{(k)}(\boldsymbol{x},\boldsymbol{y})\,
    =\, \Big(1 - \delta_{x_{k+1},x_k+1}\Big)\,
    \delta_{\boldsymbol{y},\boldsymbol{x}}\,
    - \frac{1}{2\Delta}\,
    \Big(\delta_{\boldsymbol{y},\boldsymbol{x}-\boldsymbol{e}_{k+1}}
    + \delta_{\boldsymbol{y},\boldsymbol{x}+\boldsymbol{e}_{k}}\Big)\, ;
\end{equation*}
and
\begin{equation*}
    K_n^{(n)}(\boldsymbol{x},\boldsymbol{y})\,
    =\, \frac{1}{2}\, \delta_{\boldsymbol{y},\boldsymbol{x}}
    - \frac{1}{2\Delta}\, \delta_{\boldsymbol{y},\boldsymbol{x}+
    \boldsymbol{e}_n}\, ,
\end{equation*}
This kernel is symmetric, corresponding to the fact that $H_{\Z}(n)$ 
is self-adjoint. 
The kernel can also be used to define operators
on $\ell^p(\mathcal{X}_n)$ for $p$ other than $2$. One important
preliminary step is to observe that $H_{\Z,n}$ is bounded.

\begin{lemma}
$ \displaystyle
    \|H_{\Z}(n)\|\, \leq\, n \left(1 + \Delta^{-1}\right)$.
\end{lemma}

\begin{proof}
Let $H_{\Z}(n;p)$ be the operator on $\ell^p(\mathcal{X}_n)$ with
the kernel $K_n$, as above. So $\H_{\Z}(n) = \H_{\Z}(n;2)$. 
One knows that
\begin{equation*}
    \|H_{\Z}(n;\infty)\|\,
    =\, \max_{\boldsymbol{x} \in \mathcal{X}_n}\, \sum_{\boldsymbol{y} \in
    \mathcal{X}_n} |K_n(\boldsymbol{x},
    \boldsymbol{y})|\, ,
\end{equation*}
and
\begin{equation*}
    \|H_{\Z}(n;1)\|\,
    \leq\, \max_{\boldsymbol{y} \in \mathcal{X}_n}\, \sum_{\boldsymbol{x}
    \in \mathcal{X}_n}
    |K_n(\boldsymbol{x},\boldsymbol{y})|\, .
\end{equation*}
Since the kernel is symmetric, these two numbers -- the maximum
column sum and maximum row sum -- are equal. Both are bounded by
$n ( 1 + \Delta^{-1})$. (They are actually equal to it.) This
follows by considering the maximum number of off-diagonal entries
in any row or column, which is $2n$, as well as the maximum
diagonal entry, $n$. Both occur when $x_{i+1}>x_i+1$ for all
$i=1,\dots,n-1$. By the Riesz convexity theorem, (c.f.,
\cite{Stein}, Section 5.1), this gives the stated upper bound for
$\|H_{\Z}(n;p)\|$ for all $p \in [1,\infty]$, in particular $p=2$.
\end{proof}

\subsection{Direct integral, Fourier decomposition}\label{sec:fourier}

Let us define $\boldsymbol{d} =
\boldsymbol{e}_1+\dots+\boldsymbol{e}_n\in \Z^n$. Note that if the
coordinates of $\boldsymbol{x} \in \Z^n$ are ordered as
$x_1<\dots<x_n$, then the same is true for $\boldsymbol{x} +
\boldsymbol{d}$. Therefore, the function $\tau(\boldsymbol{x}) =
\boldsymbol{x} + \boldsymbol{d}$ defines a bijection on
$\mathcal{X}_n = \{\boldsymbol{x}\in\Z^n\, :\, x_1<\dots<x_n\} $. 
It is trivial to check that
$K_n(\boldsymbol{x},\boldsymbol{y})=K_n(\tau(\boldsymbol{x}),
\tau(\boldsymbol{y}))$
for all $\boldsymbol{x},\boldsymbol{y} \in \mathcal{X}_n$.
Therefore, defining $T : \Hil_{\Z}(n) \to \Hil_{\Z}(n)$ such that $T
\delta_{\boldsymbol{x}} = \delta_{\tau(\boldsymbol{x})}$, it
follows that $T$ and $H_{\Z}(n)$ commute.
The shift operator $T$ has absolutely continuous spectrum. The
analogue of block-diagonalizing $H_{\Z}(n)$ according to the
eigenspaces of $T$ is to make a direct-integral decomposition of
$H_{\Z}(n)$ using the usual Fourier transform with respect to $T$.
We describe this in some detail, next.

Let $\mathcal{Y}_n = \{\boldsymbol{x} \in \mathcal{X}_n : 0\leq
\boldsymbol{x} \cdot \boldsymbol{d}\leq n-1\}$. There is a natural
identification of this as the quotient space $\mathcal{X}_n/\tau$.
Namely, for every $\boldsymbol{x} \in \mathcal{X}_n$, there is a
unique $\boldsymbol{y} \in \mathcal{Y}_n$ and $k \in \Z$ such that
$\boldsymbol{x} = \tau^k(\boldsymbol{y})$. We define
$[\boldsymbol{x}]$ to be this $\boldsymbol{y}$. Let
$\ell(\mathcal{X}_n)$ be the set of all sequences on
$\mathcal{X}_n$. The operator $T$ extends naturally to this
vector space. We let $\ell_0(\mathcal{X}_n)$ be the set of
$T$-invariant sequences $f \in \ell(\mathcal{X}_n)$. Moreover, we
define $\ell_0^2(\mathcal{X}_n)$ to be the Hilbert space of
functions in $\ell_0(\mathcal{X}_n)$ such that the following norm is finite:
\begin{equation*}
    \|f\|^2\, =\, \sum_{\boldsymbol{y} \in \mathcal{Y}_n}
    |f(\boldsymbol{y})|^2\, .
\end{equation*}

Let $\S^1=\R/2\pi\Z$ be the unit circle. 
Let $L^2(\S^1,\ell^2_0(\mathcal{X}_n))$ be the Hilbert space 
consisting of the set of
functions $\Phi : \S^1 \to \ell^2_0(\mathcal{X}_n)$ such that for
each $\boldsymbol{x} \in \mathcal{X}_n$ the function
$\Phi(\cdot)(\boldsymbol{x})$ is measurable on $\S^1$. The norm is
\begin{equation*}
    \|\Phi\|^2\, =\, \int_0^{2\pi} \|\Phi(\theta)\|^2\,
    \frac{d\theta}{2\pi}\, ,
\end{equation*}
where $\Phi(\theta) \in \ell^2_0(\mathcal{X}_n)$ so $\|\Phi(\theta)\|$ is 
the norm in $ \ell^2_0(\mathcal{X}_n)$.
There is an analogous Banach
space $L^p(\S^1,\ell_0^2(\mathcal{X}_n))$ for each $p \in
[1,\infty]$ such that
\begin{equation*}
    \|\Phi\|_p^p\, =\, \int_0^{2\pi} \|\Phi(\theta)\|^p\,
    \frac{d\theta}{2\pi}\, ,
\end{equation*}
where $\|\Phi(\theta)\|$ is still the norm  in $ \ell^2_0(\mathcal{X}_n)$. 
We continue to denote $\|\Phi\|_2$ by just $\|\Phi\|$. Then
$L^\infty(\S^1,\ell_0^2(\mathcal{X}_n))$ is a dense subspace of
$L^2(\S^1,\ell_0^2(\mathcal{X}_n))$, as in the case of finite-dimensional 
vector-valued functions. 

We define a map, $\mathcal{G}$, which is an analogue of the Fourier series,
by
\begin{equation*}
\mathcal{G}
: L^\infty(\S^1,\ell_0^2(\mathcal{X}_n)) \to
\ell^2(\mathcal{X}_n):
    \mathcal{G} \Phi(\boldsymbol{x})\, =\, \int_0^{2\pi}
    e^{i\theta \boldsymbol{x} \cdot \boldsymbol{d}}\,
    \Phi(\theta)(\boldsymbol{x})\, \frac{d\theta}{2\pi}\, .
\end{equation*}
As in the case of the usual Fourier series, we define the map on a dense
subspace first, and will eventually extend to the full Hilbert space using the
isometry property. (We remind the reader to think of $\ell_0^2(\mathcal{X}_n)$
as functions on $\ell^2(\mathcal{Y}_n)$, where $\mathcal{X}_n \cong
\mathcal{Y}_n \times \Z$ to see the analogy with the Fourier series.) One can
determine that for these functions,
\begin{equation*}
    \|\mathcal{G} \Phi\|^2\, =\, \frac{1}{n} \sum_{k=0}^{n-1} \int_0^{2\pi}
    \sum_{\boldsymbol{y} \in \mathcal{Y}_n}
    \Phi(\theta)(\boldsymbol{y})\,
    \overline{\Phi(\theta+2\pi k/n)(\boldsymbol{y})}\, e^{-2 \pi i k
    \boldsymbol{y} \cdot \boldsymbol{d}/n}\,
    \frac{d\theta}{2\pi}\, .
\end{equation*}
This map has a nontrivial null-space  (except in the case $n=1$) because of the
sum over $k$. But it is if we restrict to the closed subspace
$L^2_*(\S^1,\ell_0^2(\mathcal{X}_n))$ of $L^2(\S^1,\ell_0^2(\mathcal{X}_n))$, 
defined as all those $\Phi$ such that
\begin{equation}
\label{constraint}
    \Phi(\theta+2\pi/n)(\boldsymbol{x})\, =\, e^{-2 \pi i
    \boldsymbol{x} \cdot \boldsymbol{d}/n}\,
    \Phi(\theta)(\boldsymbol{x})\, ,
\end{equation}
then $\mathcal{G}$ is a partial isometry between
$L^2_*(\S^1,\ell_0^2(\mathcal{X}_n))\cap
L^{\infty}(\S^1,\ell_0^2(\mathcal{X}_n))$ (with the
$L^2(\S^1,\ell_0^2(\mathcal{X}_n))$ norm) and its range in
$L^2(\mathcal{X}_n)$. One can then extend $\mathcal{G}$ to all of
$L^2_*(\S^1,\ell_0^2(\mathcal{X}_n))$, to obtain an isometry with
$\ell^2(\mathcal{X}_n)$. 

This map is surjective. For example,
given any $\widetilde{\boldsymbol{x}} \in \mathcal{X}$, defining
\begin{equation*}
    \Phi(\theta)(\boldsymbol{x})\, =\,
    \delta_{[\boldsymbol{x}],[\widetilde{\boldsymbol{x}}]}\, e^{-i
    \theta \widetilde{\boldsymbol{x}} \cdot \boldsymbol{d}}\, ,
\end{equation*}
one can easily check that $\mathcal{G} \Phi =
\delta_{\widetilde{\boldsymbol{x}}}$. Recall that $\ell^2(\mathcal{X}_n)$ is 
$\Hil_{\Z}(n)$. Let us define the map $\mathcal{F} :
\Hil_{\Z}(n) \to L^2_*(\S^1,\ell_0^2(\mathcal{X}_n))$ as the
inverse of $\mathcal{G}$.

In a situation such as this, it is usual to call the Fourier-type
decomposition a direct-integral decomposition, and to write
$$
  \Hil_{\Z}(n)\, \overset{\mathcal{F}}{\cong}\, \int_{\S^1}^{\oplus} 
  \ell^2_0(\mathcal{X}_n)\, \frac{d\theta}{2\pi}\, .
$$
(This matches the notation of \cite{Reed1978}, section 13.16.)  Actually, we
have a slightly more involved situation because of the constraint
(\ref{constraint}). We could reduce to the usual situation by restricting
attention to $\theta \in [0,2\pi/n)$, which is a fundamental domain for such
$\Phi$. However, for notational purposes which arise shortly, we prefer to keep
the present convention, and merely remember that (\ref{constraint}) must be
satisfied.

We can define a family of bounded operators $H_{\Z}(n;\theta)$ on 
$\ell_0^2(\mathcal{X}_n)$, such that for every $f \in
\Hil_{\Z}(n)$,
\begin{equation*}
    \mathcal{F}\, H_{\Z}(n)\, f(\theta)\, =\,  H_{\Z}(n;\theta)\,
    \mathcal{F}\, f(\theta)\, ,
\end{equation*}
for all $\theta \in \S^1$. This can be done precisely because $H_{\Z}(n)$ 
commutes with $T$. One usually then says that $H_{\Z}(n)$
is a decomposable operator, and writes
$$
  H_{\Z}(n)\, \overset{\mathcal{F}}{\cong}\, \int_{\S^1}^{\oplus} 
  H_{\Z}(n;\theta)\, \frac{d\theta}{2\pi}\, .
$$
We have the same caveat about remembering (\ref{constraint}) as before. 
Particularly, we should pay attention to the fact that the proposed 
$H_{Z}(n;\theta)$ should preserve this property.

The operators $H_{\Z}(n;\theta)$ are easiest to express as discrete integral 
operators, as was the case for $H_{\Z}(n)$ itself.
For each $\theta \in \S^1$, we define
\begin{equation}
    K_{n,\theta}(\boldsymbol{x},\boldsymbol{y})\,
    =\, \sum_{k=0}^n K_{n,\theta}^{(k)}(\boldsymbol{x},\boldsymbol{y})\, ,
\end{equation}
where:
\begin{equation*}
    K_{n,\theta}^{(0)}(\boldsymbol{x},\boldsymbol{y})\,
    =\, \frac{1}{2}\, \delta_{\boldsymbol{y},\boldsymbol{x}}
    - \frac{1}{2\Delta}\, e^{i\theta}\, \delta_{\boldsymbol{y},
    \boldsymbol{x}-\boldsymbol{e}_1}\, ;
\end{equation*}
for $k=1,\dots,n-1$,
\begin{equation*}
    K_{n,\theta}^{(k)}(\boldsymbol{x},\boldsymbol{y})\,
    =\, \Big(1 - \delta_{x_{k+1},x_k+1}\Big)\,
    \delta_{\boldsymbol{y},\boldsymbol{x}}\,
    - \frac{1}{2\Delta}\,
    \Big(e^{i\theta}\, \delta_{\boldsymbol{y},\boldsymbol{x}-
    \boldsymbol{e}_{k+1}}
    + e^{-i\theta}\, \delta_{\boldsymbol{y},\boldsymbol{x}+
    \boldsymbol{e}_{k}}\Big)\, ;
\end{equation*}
and
\begin{equation*}
    K_{n,\theta}^{(n)}(\boldsymbol{x},\boldsymbol{y})\,
    =\, \frac{1}{2}\, \delta_{\boldsymbol{y},\boldsymbol{x}}
    - \frac{1}{2\Delta}\, e^{-i\theta}\, \delta_{\boldsymbol{y},\boldsymbol{x}
    +\boldsymbol{e}_n}\, ,
\end{equation*}
In addition to $H_{Z}(n;\theta)$ it is useful to consider the operator on the
vector space $\ell(\mathcal{X}_n)$ defined through this same kernel. We denote
this operator as $\mathcal{K}(\theta)$. It is easy to check that the kernel
does map the vector space to itself (i.e., there are no divergences, because
all sums are finite). It is also easy to see that $\ell_0(\mathcal{X}_n)$ is an
invariant subspace. The restriction of $\mathcal{K}(\theta)$ to
$\ell_0^2(\mathcal{X}_n)$ is $H_{\Z}(n;\theta)$.

From the decomposition, it is clear that for any $f \in \Hil_{Z}(n)$,
we have
\begin{equation*}
    \ip{f}{H_{Z}(n)\, f}\, =\, \int_0^{2\pi} \ip{\mathcal{F} f(\theta)}
    {H_{\Z}(n,\theta)\, \mathcal{F} f(\theta)}\,
    \frac{d\theta}{2\pi}\, .
\end{equation*}
Also, the map  $\theta \mapsto H_{\Z}(n;\theta)$ is norm
continuous. Therefore, by the Rayleigh-Ritz variational principle,
\begin{equation*}
    \infspec\, H_{\Z}(n)\, =\, \min_{\theta \in \S^1}\, 
    \infspec\, H_{\Z}(n;\theta)\,
    .
\end{equation*}
(At this point, the reader may be concerned that because of (\ref{constraint})
this might not be true. We leave it as an easy exercise to check that
$\infspec\, H_{\Z}(n;\theta) = \infspec\, H_{\Z}(n;\theta+2\pi k/n)$ for all
$\theta$ and that moreover, the equation above is true.) We can see that the
minimum is attained for $\theta=0$ (as well as possibly other values as per the
last comment). This follows from the observation that for any $f \in
\ell^2_0(\mathcal{X}_n)$, one has
\begin{equation*}
    \ip{|f|}{H_{\Z}(n,0)\, |f|}\, \leq\, \ip{f}{H_{\Z}(n,\theta)\, f}\, ,
\end{equation*}
for all $\theta \in \S^1$, where $|f| \in \ell^2_0(\mathcal{X}_n)$ is the
function $|f|(\boldsymbol{x})=|f(\boldsymbol{x})|$.  (This, in turn, follows
because the kernel $K_{n,0}$ has nonpositive signs for the off-diagonal
entries.) From this, one knows that Proposition \ref{prop:infspec} will follow
if we show that
\begin{equation*}
  \infspec\, H_{\Z,n}(0)\, =\, \frac{(1-q^2)(1-q^n)}{(1+q^2)(1+q^n)}\, .
\end{equation*}

\subsection{The Bethe ansatz}

We will now develop the simplest possible application of the Bethe
ansatz. The following is a well-known result, which we include for completeness.

\begin{lemma}\label{lem:BA}
Let $\C^\times  = \C \setminus \{0\}$. Suppose that
$\boldsymbol{\xi} \in (\C^\times)^n$ satisfies, for
$k=1,\dots,n-1$,
\begin{equation}
\label{eq:lfrr}
    e^{i\theta} \xi_k + e^{-i\theta} \xi_{k+1}^{-1} = 2\Delta\, .
\end{equation}
Define the function, $f_{\boldsymbol{\xi}} \in \ell(\mathcal{X}_n)$ as
\begin{equation}
\label{eq:fdef}
    f_{\boldsymbol{\xi}}(\boldsymbol{x})\, =\, \prod_{k=1}^n
    \xi_k^{x_k}\, .
\end{equation}
Then $f$ is an eigenvector of $\mathcal{K}(\theta)$ with
eigenvalue equal to
\begin{equation}
\label{eq:endef}
    E(\boldsymbol{\xi},\theta)\, =\, \sum_{k=1}^n \left(1 - \frac{1}{2\Delta}
\left[e^{i\theta}\, \xi_k + e^{-i\theta}\, \xi_k^{-1}\right]
\right)\, .
\end{equation}
\end{lemma}

\begin{proof} The key is to consider an operator on $\ell(\Z^n)$
which restricts to $\mathcal{K}(\theta)$ on $\ell(\mathcal{X}_n)$. 
Define the following kernel on $\Z^n$,
\begin{equation*}
    L_{n,\theta}(\boldsymbol{x},\boldsymbol{y})\,
    =\, \sum_{k=1}^n \Big(\delta_{\boldsymbol{y},\boldsymbol{x}} -
    \frac{1}{2\Delta}
    \left(e^{i\theta} \delta_{\boldsymbol{y},\boldsymbol{x}- \boldsymbol{e}_k}
    + e^{-i\theta} \delta_{\boldsymbol{y},\boldsymbol{x}+\boldsymbol{e}_k} 
    \right)
    \Big)\, .
\end{equation*}
Let $\mathcal{L}(\theta)$ be the operator with this kernel. This
is easily related to the Laplacian on $\Z^n$. In particular, if
one defines $F_{\boldsymbol{\xi}} \in \ell(\Z^n)$ by the same
formula as in (\ref{eq:fdef}) (except on all of $\Z^n$), one has
\begin{equation*}
    \mathcal{L}(\theta) F_{\boldsymbol{\xi}}\, =\, E(\boldsymbol{\xi},\theta)
    F_{\boldsymbol{\xi}}\, .
\end{equation*}
For $k=1,\dots,n-1$, define a kernel on $\Z^n$ by
\begin{equation*}
    M_{n,\theta}^{k,k+1}(\boldsymbol{x},\boldsymbol{y})\, =\,
    \delta_{y_{k+1},y_k+1}\,  \Big(\delta_{\boldsymbol{y},\boldsymbol{x}} -
    \frac{1}{2\Delta}
    \left(e^{i\theta} \delta_{\boldsymbol{y},\boldsymbol{x}-\boldsymbol{e}_{k}}
    + e^{-i\theta} \delta_{\boldsymbol{y},\boldsymbol{x}+\boldsymbol{e}_{k+1}} 
    \right)
    \Big)\, .
\end{equation*}
Let $\mathcal{M}_{k,k+1}(\theta)$ be the operator with this
kernel. Then we claim the following is true. First, for any
$\boldsymbol{x}, \boldsymbol{y} \in \mathcal{X}_n$,
\begin{equation*}
    K_{n,\theta}(\boldsymbol{x},\boldsymbol{y})\, =\,
    L_{n,\theta}(\boldsymbol{x},\boldsymbol{y}) - \sum_{k=1}^{n-1}
    M_{n,\theta}^{k,k+1}(\boldsymbol{x},\boldsymbol{y})\, .
\end{equation*}
Second, if $\boldsymbol{y} \in \mathcal{X}_n$ and $\boldsymbol{x}
\in \Z^n \setminus \mathcal{X}_n$, then
\begin{equation*}
    L_{n,\theta}(\boldsymbol{x},\boldsymbol{y}) - \sum_{k=1}^{n-1}
    M_{n,\theta}^{k,k+1}(\boldsymbol{x},\boldsymbol{y})\,
    =\, 0\, .
\end{equation*}
The reader can check both properties easily. Because of these two
properties the following is a fact. Suppose that $F \in
\ell(\Z^n)$ is an eigenvector of $\mathcal{L}(\theta)$ with
eigenvalue $E(\theta)$, and suppose that
$\mathcal{M}_{k,k+1}(\theta) F=0$ for all $k=1,\dots,n-1$. Then
defining $f \in \ell(\mathcal{X}_n)$ to be the restriction of $F$,
one knows that $f$ is an eigenvector of $\mathcal{K}(\theta)$ with the same
eigenvalue $E(\theta)$.

The condition
$M_{k,k+1}(\theta)
F_{\boldsymbol{\xi}}=0$, is called the ``meeting condition'' in the context 
of the Bethe ansatz. It is
\begin{equation*}
    2\Delta\, F(\boldsymbol{y})\, =\,
    e^{i\theta} F(\boldsymbol{y}+\boldsymbol{e}_k)
    + e^{-i\theta} F(\boldsymbol{y}-\boldsymbol{e}_{k+1})\, ,
\end{equation*}
for every $\boldsymbol{y} \in \Z^n$ such that $y_{k+1}=y_k+1$. For
$F = F_{\boldsymbol{\xi}}$, this is equivalent to
\begin{equation*}
    2\Delta\, \xi_k^{y_k}\, \xi_{k+1}^{y_k+1}\, =\,
    e^{i\theta} \xi_k^{y_k+1}\, \xi_{k+1}^{y_k+1} + e^{-i\theta} \xi_k^{y_k}\,
    \xi_{k+1}^{y_k}\, .
\end{equation*}
Dividing by $\xi_k^{y_k}\, \xi_{k+1}^{y_k+1}$, this is precisely
the relation in (\ref{eq:lfrr}).
\end{proof}

The condition in (\ref{eq:lfrr}) is the same as the linear
fractional relation
\begin{equation*}
    e^{i\theta} \xi_{k+1}\, =\, \frac{1}{2\Delta - e^{i\theta} \xi_k}\, .
\end{equation*}
Let us define the matrix
\begin{equation*}
    A\, =\, \begin{bmatrix} 0 & 1 \\ -1 & q+q^{-1} \end{bmatrix}\, .
\end{equation*}
Then the linear fractional relation is expressed as
\begin{equation*}
    A\, \begin{bmatrix} e^{i\theta} \xi_k \\ 1 \end{bmatrix}\,
    =\, \begin{bmatrix} 1 \\ e^{-i\theta} \xi_{k+1}^{-1} \end{bmatrix}\, .
\end{equation*}
More generally, suppose $v_k \in \C^2\setminus\{0\}$ is a vector
such that $v_k^1/v_k^2 = e^{i\theta} \xi_k$. Then, defining
$v_{k+1} = A v_k$, we see that $v_{k+1} \in \C^2\setminus\{0\}$
and $v_{k+1}^1/v_{k+1}^2 = e^{i\theta} \xi_{k+1}$. The eigenpairs
of $A$ are
\begin{equation*}
    \lambda_+\, =\, q\, ,\quad v_+\, =\,
    \begin{bmatrix}q^{-1/2}\\q^{1/2}\end{bmatrix}\qquad \textrm{and}
    \qquad \lambda_-\, =\, q^{-1}\, ,\quad v_-\, =\,
    \begin{bmatrix}q^{1/2}\\q^{-1/2}\end{bmatrix}\, .
\end{equation*}
Therefore, the most general solution to the linear recurrence
relation $v_{k+1}=A v_k$ for $k=1,\dots,n-1$ is
\begin{equation*}
    v_k\, =\, \alpha\, q^{k}\, v_+ + \beta\, q^{-k}\, v_-\, .
\end{equation*}
Therefore, the most general solution to the linear fractional
recurrence relation (\ref{eq:lfrr}) is
\begin{equation*}
    e^{i\theta} \xi_k\, =\, \frac{z^{1/2}\, q^{k-1/2} + z^{-1/2}\, q^{-k+1/2}}
    {z^{1/2}\, q^{k+1/2} + z^{-1/2}\, q^{-k-1/2}}\, .
\end{equation*}
(For this, we have taken $\alpha=z^{1/2} q^{1/2}$ and $\beta = z^{-1/2} 
q^{-1/2}$, 
which is allowed since the two variables $\alpha$ and $\beta$ only amount 
to one 
independent quantity in the ratio.) Let us define
\begin{equation*}
    \Xi_{m}(z)\, =\, \frac{z^{1/2}\, q^{m-1/2} + z^{-1/2}\, q^{-m+1/2}}
    {z^{1/2}\, q^{m+1/2} + z^{-1/2}\, q^{-m-1/2}}\, ,
\end{equation*}
for all $z$. Then another way to write the most general solution
of (\ref{eq:lfrr}) is $\xi_k\, =\, \Xi_{m(k)}(z)$, where
$m(k)=k-(n+1)/2$ and $z \in \C$. If we wish to have
$T$-invariance, then we require $\xi_1 \cdots \xi_n=1$. Since it
is more convenient to work with $e^{i\theta} \xi_k$, this is rewritten
as $(e^{i\theta} \xi_1)\cdots (e^{i\theta}
\xi_n)=e^{in\theta}$. One easily sees that
\begin{equation*}
    \Xi_{-M}(z) \Xi_{-M+1}(z) \cdots \Xi_{M}(z)\,
    =\, \frac{z^{1/2}\, q^{-M-1/2} + z^{-1/2}\, q^{M+1/2}}
    {z^{1/2}\, q^{M+1/2} + z^{-1/2}\, q^{-M-1/2}}\, .
\end{equation*}
So the condition for $T$-invariance is
that $z$ solves
\begin{equation*}
    \frac{z^{1/2}\, q^{-M-1/2} + z^{-1/2}\, q^{M+1/2}}
    {z^{1/2}\, q^{M+1/2} + z^{-1/2}\, q^{-M-1/2}}\,
    =\, e^{in\theta}\, ,
\end{equation*}
where $M=(n-1)/2$. One solution is $z=e^{i\Theta}$, where
$\Theta=\Theta(q,n,\theta)$ is
\begin{equation*}
    \Theta\, =\, 2
    \tan^{-1}\left(\frac{1+q^n}{1-q^n}\tan(n\theta/2)\right)\, .
\end{equation*}

For each choice of $N_2,\dots,N_n \in
\N=\{1,2,3,\dots\}$, there is a unique point $\boldsymbol{y} \in 
\mathcal{Y}_n$ with
$y_k-y_{k-1}=N_k$ for $k=2,\dots,N$, and this labels all possible
points in $\mathcal{Y}_n$. Therefore, one has
\begin{equation*}
    \sum_{\boldsymbol{y} \in \mathcal{Y}_n} 
    |f_{\boldsymbol{\xi}}(\boldsymbol{y})|^2\,
    =\, \sum_{N_2,\dots,N_n \in \N}\, \prod_{k=2}^n \Bigg(\prod_{j=k}^n 
    |\xi_j|\Bigg)^{N_k}\,
    =\, \prod_{k=2}^n \left(\frac{\prod_{j=k}^n |\xi_j|}{1 - \prod_{j=k}^n
    |\xi_j|}\right)\, .
\end{equation*}
But, for $2\leq k\leq n$, we have
\begin{eqnarray*}
    \prod_{j=k}^n \xi_j\,
    &=&\, \Xi_m(z)\cdots \Xi_M(z)\,\\
    &=&\, \frac{\cos(\Theta/2) [q^{m-1/2} + q^{-m+1/2}] + i
    \sin(\Theta/2) [q^{m-1/2} - q^{-m+1/2}]}{\cos(\Theta/2)
    [q^{M+1/2} + q^{-M-1/2}] + i \sin(\Theta/2)
    [q^{M+1/2}-q^{-M+1/2}]}\, ,
\end{eqnarray*}
for $M=(n-1)/2$ and $m=(2k-n-1)/2$. This clearly has norm less than 1. So $\|f_{\boldsymbol{\xi}}\|$, relative to $\ell_0^2(\mathcal{X}_n)$, is finite.

The eigenvalue is
\begin{equation*}
    E(\boldsymbol{\xi},\theta)\, =\, \sum_{m=-M}^M \left( 1 -
    \frac{1}{2\Delta}[\Xi_m(z) + \Xi_m(z)^{-1}]\right)\, ,
\end{equation*}
where $z=e^{i\Theta}$. But one can easily verify that
\begin{equation*}
    1 -
    \frac{1}{2\Delta}[\Xi_m(z) + \Xi_m(z)^{-1}]\,
    =\, \frac{1-q^2}{1+q^2} \left[\frac{1}{1+q^{2m+1}z} -
    \frac{1}{1+q^{2m-1}z}\right]\, .
\end{equation*}
Therefore, by a telescoping sum
\begin{eqnarray*}
    E(\boldsymbol{\xi},\theta)
    &=&
    \frac{1-q^2}{1+q^2} \left[\frac{1}{1+q^{2M+1}z} -
    \frac{1}{1+q^{-2M-1}z}\right]\\
    &=& \frac{1-q^2}{1+q^2} \left[\frac{1}{1+q^{n}z} -
    \frac{1}{1+q^{-n}z}\right]\, .
\end{eqnarray*}
Putting this all together, we obtain the following.
\begin{lemma}
\label{lem:Bethe} Given $n \in \N_+$, we set $M=(n-1)/2$. For 
$\theta \in (-\pi/n,\pi/n)$, we define
\begin{equation*}
    \Theta\, =\, 2
    \tan^{-1}\left(\frac{1+q^n}{1-q^n}\tan(n\theta/2)\right) \in (-\pi,\pi)\, ,
\end{equation*}
and for $m=-M,-M+1,\dots,M$ we define
\begin{equation*}
    \Xi_m\, =\, \frac{q^{m-1/2} e^{i\Theta/2} + q^{-m+1/2}
    e^{-i\Theta/2}}{q^{m+1/2} e^{i\Theta/2} + q^{-m-1/2}
    e^{-i\Theta/2}}\, .
\end{equation*}
Then setting
\begin{equation*}
    \xi_k\, =\, e^{-i\theta}\, \Xi_{k-(n+1)/2}\, ,
\end{equation*}
for $k=1,\dots,n$, we have that
$f_{\boldsymbol{\xi}}(\boldsymbol{x})=\prod_{k=1}^n \xi_k^{x_k}$
defines a (normalizable) eigenvector of $H_{\Z}(n;\theta)$. Its energy
eigenvalue is equal to
\begin{equation*}
    E_n(\theta)\, =\,
    \frac{(1-q^2)(1-q^{2n})}{(1+q^2)(1+q^{n} e^{i\Theta})(1+q^n e^{-i\Theta})}
    \, .
\end{equation*}
In particular, when $\theta=0$, this gives the formula from
(\ref{eq:form}).
\end{lemma}

\begin{remark}
Parts of this lemma are standard. For example, the linear
fractional transformation was solved by Babbitt and Gutkin in
\cite{Babbitt1990}. However, they did not consider the direct
integral decomposition. Instead they considered ``generalized
eigenvectors''. Also, no reference is made to the exact formula
for the energy. A more explicit formula is 
$$ E_n(\theta)\, =\,\frac{1-q^2}{(1+q^2)(1+q^{n})} \left(
   1-q^n + \frac{2(1-\cos{\theta})}{1-q^n}\right) \,.
$$   
From this we can derive the well-known dispersion relation for the  isotropic
model, $\lim_{q\uparrow1} E_n(\theta) = \frac{1}{n} (1-\cos\theta)$. However,
to prove that this is the minimum energy for  $H_{\Z}(n;\theta)$ is beyond our
calculations because we would have to obtain the full diagonalization of
$H_{\Z}(n;\theta)$, which we have not done. For $q=1$, this was done in the
important work by Babbitt and Thomas~\cite{Babbitt1977b}. (The generalization
to other $q$ was outlined in \cite{Babbitt1990}.) All that we need to calculate
is $E_n(\theta=0)$, which we handle by a different technique, next.
\end{remark}

\begin{remark}
In \cite{Yang1966c}, Yang and Yang considered the ferromagnetic XXZ
model to complement their famous and important work on the
antiferromagnetic XXZ model \cite{Yang1966a}. They derived the linear 
fractional
recurrence relation for $\theta=0$ with respect to the problem of
calculating $E^{\per}(n)$. However, they did not solve the linear
fractional recurrence relation, although they did set up a
graphical method of solution which allowed them to determine the
important fact that $\lim_{m \to +\infty} \Xi_m = q$ and $\lim_{m
\to -\infty} \Xi_m = 1/q$.

This would have given them the result that $\lim_{n \to \infty}
E^{\per}(n)$ is finite. (In fact it is 
$\alpha=(1-q^2)/(1+q^2)$.) But there is an unfortunate
typographical error in their paper. They mistyped the formula for
the energy in the equation just before equation (24) of \cite{Yang1966c}
(compare to their definition in equation (11) of \cite{Yang1966a}). 
In our notation, their error is equivalent to saying
$E_n(\boldsymbol{\xi},\theta=0)$ is equal to
\begin{equation*}
    \sum_{k=1}^n \left[1 - (\xi_k + \xi_k^{-1})\right]\, .
\end{equation*}
In other words, they left off an important factor
$(2\Delta)^{-1}$. For this reason, they determined that the energy of
an ``edge spin'' is asymptotically equal to $-(2\Delta-1)$ (with
our notation) instead of the correct value,
which is 0. As a consequence the droplet nature of these states
was not recognized at the time.
\end{remark}

\subsection{Positive eigenvectors are ground states}

In Lemma \ref{lem:Bethe}, setting $\theta=0$ gives $\Theta=0$ and
therefore $\xi_k>0$ for all $k$. Hence $f_{\boldsymbol{\xi}}$ is a
strictly positive eigenvector of $H_{\Z}(n;0)$. Moreover, we claim
that $\infspec H_{\Z}(n;0)$ is the eigenvalue of
$f_{\boldsymbol{\xi}}$. This would be enough to prove Proposition
\ref{prop:infspec}. We also know that $c \boldsymbol{1} -
H_{\Z}(n;0)$ is positivity preserving (when considered as an
operator on $\ell^2(\mathcal{Y}_n)$) and bounded. Therefore, the
proof is completed by applying the following theorem.

\begin{thm}\label{thm:PF}
Let $\mathcal{Y}$ be a countable set. Suppose that $A$ is a
positivity preserving, bounded, self-adjoint operator on
$\ell^2(\mathcal{Y})$. If $A$ has a strictly positive eigenvector
$f$, then the eigenvalue of $f$ equals the spectral radius.
\end{thm}

\begin{proof}
Without loss of generality, assume $\mathcal{Y} = \N_+$.  Let us denote 
$\Hil=\ell^2(\N_+)$. For each
$N \in \N_+$, let $P_N$ be the orthogonal projection from $\Hil$ onto
$\Hil_N=\ell^2(\{1,\dots,N\})$. Let us define
$$
  A_N\, :=\, P_N\, A\, P_N\, .
$$
Obviously this is positivity preserving.
Let us
also define
$$
  f_N\, :=\, P_N\, f\, ,
$$
which is a strictly positive vector in $\Hil_N$.
Let $E$ be the eigenvalue of $f$. Then
\begin{equation}
\label{littleformula}
  P_N\, A\, f\, =\, E\, f_N\, .
\end{equation}
Let us define $P_N'=\unity-P_N$, and let us define another
nonnegative vector in $\Hil_N$,
$$
  \widetilde{f}_N\, :=\, P_N\, A\, P_N'\, f\, .
$$
Since $P_N'f = f - f_N$, an obvious bound is $\|\widetilde{f}_N\|
\leq \|A\| \cdot \|f - f_N\|$.
Using (\ref{littleformula}), we have
\begin{equation}
\label{approxeig}
 A_N\, f_N + \widetilde{f}_N\, =\, E\, f_N\, .
\end{equation}

Let $t_N  = \big(f_N,\widetilde{f}_N\big)\geq 0$ and $r_N =
\|f_N\|^2>0$. We know that $r_N$ is an increasing sequence with
limit equal to $\|f\|^2$. Consider the operator $\widetilde{A}_N$,
which is self-adjoint on $\Hil_N$, defined by
$$
  \widetilde{A}_N\, g =\, A_N\, g + r_N^{-1}\, \left[\widetilde{f}_N\, 
  \ip{f_N}{g}
  + f_N\, \big(\widetilde{f}_N,g\big)\right]\, .
$$
Note that each of the three summands is positivity preserving,
since $f_N$ and $\widetilde{f}_N$ are nonnegative vectors. On the
other hand,
$$
  \widetilde{A}_N\, f_N\, =\, A_N\, f_N + \widetilde{f}_N + 
  \frac{t_N}{r_N}\, f_N\, .
$$
Therefore, by equation (\ref{approxeig}), we have that
$$
  \widetilde{A}_N\, f_N\, =\, \left(E + \frac{t_N}{r_N}\right)\, f_N\, .
$$
In other words, $f_N$ is an eigenvector of $\widetilde{A}_N$. Since 
$f_N$ has strictly positive components in
$\Hil_N$, the Perron-Frobenius theorem guarantees
that the spectral radius of $\widetilde{A}_N$ equals $E +
(t_N/r_N)$. From this we determine that
$$
  \lim_{N \to \infty} \max\spec(\widetilde{A}_N)\, =\, E\, ,
$$
because $r_N \uparrow \|f\|^2$, while
$$
  0\leq t_N \leq \|f_N\|\cdot \|\widetilde{f}_N\|\, \leq\, \|A\|\cdot 
  \|f\|\cdot \|f-f_N\|\, ,
$$
and $\|f-f_N\| \downarrow 0$ as $N \to \infty$.

We claim that $\rho(A) \leq   \lim_{N \to \infty}
\rho(\widetilde{A}_N)$ by the variational principle. This would
imply that $\rho(A) \leq E$. We already know that $E$ is in the
spectrum of $A$, so that if $\rho(A) \leq E$ we have $\rho(A)=E$.
So we just need to prove $\rho(A)\leq \lim_{N \to \infty}
\rho(\widetilde{A}_N)$.

To begin with, note that by the (Rayleigh-Ritz) variational
principle for self-adjoint operators, we have
\begin{equation}
\label{varprin}
  \rho(A)\, =\, \sup\{ \ip{g}{A\, g}\, :\, \|g\|=1\}\, .
\end{equation}
Therefore, it suffices to observe that for any $g \in \Hil$, we
have that $g=\lim_{N\to\infty} P_N g$ and 
$$
  A\, g\, =\, \lim_{N \to \infty} \widetilde{A}_N\, P_N\, g\, .
$$
The fact that $g=\lim_{N\to\infty} P_N g$ is the usual density result 
(which we have already implicitly used).
Since $\lim_{N \to \infty}  \|g - P_N\, g\|^2 = 0$, one
has
\begin{eqnarray*}
    \lim_{N \to \infty}\, \|A\, g - A_N\, P_N\, g\|^2 &=&
    \lim_{N \to \infty}\, \ip{g-P_N\, g}{A^2\, (g - P_N\, g)}\\
    &\leq& \|A\|^2\, \lim_{N \to \infty} \|g - P_N\, g\|^2\,
    =\, 0\, ,
\end{eqnarray*}
and the perturbation
$$
  [\widetilde{A}_N - A_N]\, g =\, r_N^{-1}\, \left(f_N\, 
  \ip{\widetilde{f}_N}{g}
  + \widetilde{f}_N\, \ip{f_N}{g}\right)\, ,
$$
is bounded, in norm, by $\|\widetilde{A}_N - A_N\| \leq
2\|\widetilde{f}_N\|/\|f_N\|$. This converges to zero for reasons
we have explained before. Therefore, for any $g$ satisfying
$\|g\|=1$,
\begin{equation*}
  \ip{g}{A\, g}\, =\, \lim_{N \to \infty} 
  \ip{P_N g}{\widetilde{A}_N\, P_N\, g}\,
  \leq\, \lim_{N \to \infty} \rho(A_N)\, ,
\end{equation*}
as was claimed.
\end{proof}

\section{Droplet energies in the droplet Hamiltonian and cyclic
chain}
\label{sec:dropletcyclic}

Our next goal is to compare the energies (recall their definitions in
(\ref{def:droplet}--\ref{def:Ecyc})) $E(L,n)$, $E^{\per}(L,n)$
and $E^{\delta}(L,n)$, for $\delta\geq 1$, to $E_{\Z}(n)$. The
simplest case is $E^{\delta}(L,n)$ for $\delta\geq 1$. The desired
result follows by two applications of the Rayleigh-Ritz
variational principle.

\begin{lemma}
\label{lem:droptoZ1} For any $\delta \in \R$,
\begin{equation*}
    E_{\Z}(n)\, \geq\, \limsup_{L \to \infty} E^{\delta}(L,n)\, .
\end{equation*}
\end{lemma}

\begin{proof}
Given $a\leq b$, both in $\Z$, define $\mathcal{X}_{n}([a,b]) =
\{\boldsymbol{x} \in \Z^n\, :\, a\leq x_1<\dots<x_n\leq b\}$. By definition,
\begin{equation*}
    E_{\Z}(n)\, =\, \inf\{\ip{f}{H_{\Z} f}\, :\, f \in
    \Hil_{\Z}(n)\, ,\ \|f\|=1\}\, .
\end{equation*}
Since the functions with finite support are dense in
$\Hil_{\Z}(n)$, and since $H_{\Z}$ is bounded on $\Hil_{\Z}(n)$,
one can replace this by
\begin{equation*}
    E_{\Z}(n)\, =\, \inf_{L \in \N_+}\, \inf\big\{\ip{f}{H_{\Z} f}\,
    :\, f \in \ell^2\big(\mathcal{X}_{n}([-L,L])\big)\, ,\ \|f\|=1\big\}\, .
\end{equation*}
Moreover, by translation-invariance one can shift a function on
$[-L,L]$ to a function on $[2,2L+2]$.
Therefore,
\begin{equation*}
    E_{\Z}(n)\, =\, \lim_{L \geq 3}\, \inf\big\{\ip{f}{H_{\Z} f}\,
    :\, f \in \ell^2\big(\mathcal{X}_{n}([2,L-1])\big)\, ,\ \|f\|=1\big\}\, .
\end{equation*}
Suppose $f \in \ell^2(\mathcal{X}_{[2,L-1],n}$ and $\|f\|=1$.
Define $\psi \in \Hil_{[1,L]}(n)$ by
\begin{equation*}
    \psi =\, \sum_{1\leq x_1<\dots<x_n \leq L}
    f(x_1,\dots,x_n)\, S_{x_1}^- \cdots S_{x_n}^-\,
    \ket{\Uparrow}_{[1,L]}\, .
\end{equation*}
Then it will be apparent that $\ip{\psi}{h_{x,x+1}\psi} =
\ip{f}{h_{x,x+1}f}$ for $\{x,x+1\} \subset [1,L]$. Moreover, it is
apparent that $\ip{\psi}{S^3_1 \psi} =1/2$ and that $\ip{\psi}{S^3_L
\psi} =1/2$ because $f$ vanishes if there is any down-spin at sites
$1$ or $L$. Therefore, $\ip{\psi}{H^{\delta}_{[1,L]}\psi} =
\ip{\psi}{H_{[1,L]}\psi}$ and this is equal to $\ip{f}{H_{\Z}f}$.
So
\begin{equation*}
    \ip{f}{H_{\Z} f}\, =\, \ip{\psi}{H^{\delta}_{[1,L]} \psi}\, \geq\, 
    E^{\delta}(L,n)\, .
\end{equation*}
In fact, by exactly the same argument,
\begin{equation*}
    \ip{f}{H_{\Z} f}\, \geq\, E_{L'}^{\delta}(n)\, ,
\end{equation*}
for any $L'\geq L$. From this, one sees that
\begin{equation*}
    \inf\{\ip{f}{H_{\Z} f}\, :\, f \in
    \ell^2(\mathcal{X}_{[2,L-1],n})\, ,\ \|f\|=1\}
    \geq\, \sup_{L' \geq L} E^{\delta}(L',n)\, .
\end{equation*}
But taking the limit as $L \to \infty$, we obtain the result.
\end{proof}

\begin{lemma} \label{lem:droptoZ2} For any $\delta \geq 1$, one has
\begin{equation*}
    E_{\Z}(n)\, \leq\, \liminf_{n \to \infty} E^{\delta}(L,n)\, .
\end{equation*}
\end{lemma}

\begin{proof}
Suppose that $\psi \in \Hil(L,n)$ is any normalized vector, so that 
$\|\psi\|=1$. We can write this vector as
\begin{equation*}
    \psi =\, \sum_{1\leq x_1<\dots<x_n \leq L}
    F(x_1,\dots,x_n)\, S_{x_1}^- \cdots S_{x_n}^-\,
    \ket{\Uparrow}_{[1,L]}\, ,
\end{equation*}
where $F:\mathcal{X}_{[1,L],n} \to \C$ is such that
\begin{equation*}
    \|F\|^2\, :=\, \sum_{\boldsymbol{x} \in \mathcal{X}_{[1,L],n}}
    |F(\boldsymbol{x})|^2\, =\, 1\, .
\end{equation*}
Then we can extend $F$ to a function on $\mathcal{X}_n$ by defining
\begin{equation*}
    f(\boldsymbol{x})\, =\, \begin{cases} F(\boldsymbol{x}) &
    \textrm{if } \boldsymbol{x} \in \mathcal{X}_{[1,L],n}\, ,\\
    0 & \textrm{if } \boldsymbol{x} \in \mathcal{X}_n \setminus 
    \mathcal{X}_{[1,L],n}\, .\end{cases}
\end{equation*}
It is easy to see, from the construction of the single site
representations of $\textrm{SU}(2)$ on $\Hil_{\Z}$, that
$\ip{f}{h_{x,x+1}f} = \ip{\psi}{h_{x,x+1}\psi}$ for $\{x,x+1\}
\subset [1,L]$. It is also easy to see that $\ip{f}{h_{x,x+1}f}=0$
for $\{x,x+1\} \subset \Z \setminus [1,L]$, because, in this case,
$f$ has no down-spin at either $x$ or $x+1$.
Furthermore, one can easily see that
\begin{equation*}
    \ip{f}{h_{0,1}f}\, =\, \ip{\psi}{\widetilde{h}^{+}_1
    \psi}\quad \textrm{and}\quad
    \ip{f}{h_{L,L+1}f}\, =\,
    \ip{\psi}{\widetilde{h}^{+}_L \psi}\,,
\end{equation*}
where $h^{+}_{x}\, =\, \operatorname{Tr}_{y} (h_{x,y}\, P^+_y)$ is
the partial trace of the XXZ interaction with $P^+_y = \frac{1}{2}
\unity + S_y^3$ the projection onto the up-spin vector. But one
easily calculates
\begin{equation*}
    h^{+}_x\, =\, \frac{1}{4} \unity - \frac{1}{2} S_x^{3}\, =\,
    \frac{1}{2} (\unity - P_x^+)\, .
\end{equation*}
Therefore, one sees that the energy of $f$ relative to $H_{\Z}$ is
exactly equal to
\begin{equation*}
    \ip{f}{H_{\Z} f}\, =\, \ip{\psi}{H^\delta_{[1,L]} \psi}\, ,
\end{equation*}
with $\delta=1$. Since the boundary field operator is positive
semi-definite, one obtains
\begin{equation*}
    \ip{f}{H_{\Z} f}\, \leq\, \ip{\psi}{H^\delta_{[1,L]} \psi}\, ,
\end{equation*}
as long as $\delta\geq 1$. From this, it follows that
\begin{equation*}
    E_{\Z}(n)\, \leq\, \ip{\psi}{H^\delta_{[1,L]} \psi}\, .
\end{equation*}
By minimizing over $\psi$, and using the Rayleigh-Ritz variational
principle, one obtains the result.
\end{proof}

Combining these two lemmas proves the following.

\begin{prop}
For $\delta\geq 1$, $\lim_{L \to \infty} E^{\delta}(L,n)$ exists
and equals $E_{\Z}(n)$. $\square$
\end{prop}

Let us now consider $E^{\per}(L,n)$. There is a proof completely
analogous to Lemma \ref{lem:droptoZ1} for the following.
\begin{lemma}
\label{lem:pertoZ1} $\displaystyle E_{\Z}(n)\, \geq\, \limsup_{L
\to \infty} E^{\per}(L,n)$. 
\end{lemma}

To prove the analogue of Lemma \ref{lem:droptoZ2} requires a
different argument. Given any $x \in [1,L]$, we define the
projections $P^+_x$ and $P^-_x$ on $\Hil_{[1,L]}$ by
\begin{equation*}
    P^+_x\, =\, \frac{1}{2} \unity - S_x^{3}\quad
    \textrm{and}\quad
    P^-_x\, =\, \unity - P^+_x\, .
\end{equation*}
Suppose that $[a,b] \subset [1,L]$. Then we define
\begin{equation*}
    P_{[a,b]}^{\pm}\, =\, \prod_{x=a}^b P_x^{\pm}\, .
\end{equation*}
Let us also define $P_{[a,b]} = P^+_{[a,b]} + P^-_{[a,b]}$. The
following result follows from Corollary 4.3 in \cite{Nachtergaele2001}.
\begin{lemma}
\label{lem:polarized} Let $K_L$ be a self-adjoint operator on
$\Hil_{[1,L]}$ and $M = \|K_L - H_L\|$. Given $E<\infty$, suppose
$\psi \in \Hil_{[1,L]}$ is any vector with $\|\psi\|=1$ and
$\ip{\psi}{K_L \psi}\leq E$. Given any interval $[a,a+\ell-1]
\subset [1,L]$ and any $\ell'<\ell$, there is a subinterval
$[b,b+\ell'-1] \subset [a,a+\ell-1]$ such that
\begin{equation*}
    \|P_{[b,b+\ell'-1]} \psi\|\, \geq\, 1 - \epsilon\, ,
\end{equation*}
where
\begin{equation*}
    \epsilon\, =\, \frac{2(E+M)}{\gamma \floor{\ell/\ell'}}\, .
\end{equation*}
Moreover, when $\epsilon<1$, defining $\mathcal{I}=[b,b+\ell'-1]$,
\begin{equation*}
    \ip{P_{\mathcal{I}} \psi}{K_L P_{\mathcal{I}} \psi}\,
    \leq\, \ip{\psi}{K_L \psi} + \Big(M\epsilon +
    2(\Delta^{-1}+2M)\sqrt{\epsilon(1-\epsilon)}\Big)\, .
\end{equation*}
\end{lemma}

We can now prove the following.
\begin{lemma}
$\displaystyle \liminf_{L \to \infty} E^{\per}(L,n) \geq
E_{\Z}(n)$.
\end{lemma}

\begin{proof}
The result is trivial for $n=0$. Let us assume $n>0$.

Given $L \in \N_+$, let $\psi_L$ be a vector in $\Hil_{[1,L]}(n)$
such that $\|\psi\|=1$ and $\ip{\psi_L}{H^{\per}_{L} \psi_L} =
E^{\per}(L,n)$. We will now apply Lemma \ref{lem:polarized}. We
take $\ell=L$ so that $a=1$ and $a+\ell-1=L$. We take $\ell'=2n$.
We have $M=1$ because $H_L - H^{\per}_L=h_{1,L}$ and
$\|h_{1,L}\|=1$. By Lemma \ref{lem:pertoZ1} we can take
$E=2E_{\Z}(n)$ as long as $L$ is large enough. Hereafter, we
assume that $L$ is large enough. Then we obtain
\begin{equation*}
    \epsilon\, =\, \epsilon_L\, :=\, \frac{4E_{\Z}(n) + 2}{\gamma
    \floor{L/2n}}\, .
\end{equation*}
We obviously have $\lim_{L \to \infty} \epsilon_L=0$ since
$\epsilon_L=O(1/L)$. On the other hand, given any interval
$\mathcal{I} \subset [1,L]$ with $|\mathcal{I}|=2n$, we have
\begin{equation*}
    P^{-}_{\mathcal{I}} \psi_L\, =\, 0\, ,
\end{equation*}
since $\psi_L$ only has $n$ downspins, and $P^{-}_{\mathcal{I}}$
projects onto vectors which have downspins at all $2n$ sites in
$\mathcal{I}$. Therefore, we obtain
\begin{equation*}
    \|P^{+}_{\mathcal{I}} \psi_L\|^2\, \geq\, 1-\epsilon\, .
\end{equation*}
Since we have translation invariance, we can translate the
interval $\mathcal{I}$ as long as we translate $\psi_L$ by the
same amount. Then, without loss of generality, we may assume that
\begin{equation*}
    \|P^{+}_{[1,n]}P^{+}_{[L-n+1,L]} \psi_L\|^2\, \geq\,
    1-\epsilon\, .
\end{equation*}
Let $\psi_L' = P^{+}_{[1,n]}P^{+}_{[L-n+1,L]} \psi_L$. This can be
written as
\begin{equation*}
    \psi_L'\, =\, \sum_{\boldsymbol{x} \in
    \mathcal{X}_{[n+1,L-n],n}} f_L'(\boldsymbol{x})\,
    S_{x_1}^-\cdots S_{x_n}^-\, \ket{\Uparrow}_{[1,L]}\, ,
\end{equation*}
for some $f_L' \in \ell^2(\mathcal{X}_{[n+1,L-n],n})$. It is
easily seen that $\|f_L'\|=\|\psi_L'\|$ and that $\ip{f_L'}{H_{\Z}
f_L'} = \ip{\psi_L'}{H^{\per}_{\Z} \psi_L'}$. Therefore, using the
second half of Lemma \ref{lem:polarized} and the fact that
$\lim_{L \to \infty} \epsilon_L=0$, we conclude that
\begin{equation*}
    \liminf_{L \to \infty} E^{\per}(L,n)\, \geq\, E_{\Z}(n)\, ,
\end{equation*}
as desired.
\end{proof}

Thus we have proved the following.

\begin{prop}\label{prop:cycl} For each $n \in \N$,
$\lim_{L \to \infty} E^{\per}(L,n)$ exists and equals $E_{\Z}(n)$.
$\square$
\end{prop}

\begin{remark} 
Yang and Yang actually calculated $\lim_{L \to
\infty} E^{\per}(L,n)$ using the Bethe ansatz in \cite{Yang1966c}.
However, their proof was more involved than ours, because
they used the Bethe ansatz for finite $L$ and took the asymptotic
limit. Also, as mentioned before, because of a typographical
error, they miscalculated $E^{\per}(n)$.
\end{remark}

\

\section{Droplet energies in the kink Hamiltonian}
\label{sec:Kink}

highest weight vectors are vectors with total spin equal to the total 
$S^3$-component. We denoted the subspace of highest weight vectors  on the
chain $[1,L]$ with total spin $L/2-n$ by $\Hil^\HW(n)$. We also defined  the
energy $E(L,n)$ to be the lowest energy of the kink Hamiltonian $H^{\rm k}_n$,
which is the restriction of $H^{\rm k}_{[1,L]}$ to $\Hil^\HW(n)$. We are
interested in the limit, $\lim_{L\to\infty}E(L,n)$. Throughout this section, we
require $0<q<1$; for $q=1$ we can use the $\textrm{SU}(2)$ symmetry to show 
instantly that $\lim_{L\to\infty}E(L,n)=0$. Here is the main result.

\begin{prop} \label{prop: kinkgs}
For each $n \in \N$ and $0<q<1$, 
\begin{equation} \lim_{L \to \infty} E(L,n) \,=\, E_{\Z}(n) \,= \,
\frac{(1-q^2)(1-q^n)}{(1+q^2)(1+q^n)}\,.
\end{equation}
\end{prop}

The essential part of this statement relates the ground state energy  of the
kink Hamiltonian for highest weight vectors with the ground state energy of the
translation invariant Hamiltonion in the $n$-magnon sector. A detailed
description of this relation is a bit lengthy but the first steps are
well-known and standard. First we recall the construction of a basis for the
heightest  weight vectors for the finite chain. Then, we extend this basis to
infinite volume  and define the action of the infinite volume kink Hamiltonian,
$H^{\rm k}_{\Z,n}$.  We then construct a continuous bijection, $R$, between the
highest weight  vectors and the $n$-magnon vectors with the property that
$H^{\rm k}_{\Z,n} =  R^{-1} H_{\Z,n} R$. The main proposition is proved in
Section  \ref{Sect:proof of kinkgs}.

\subsection{Basis for the highest weight vectors}

We use the notation $[2]_q=q+q^{-1}$. Let us start by introducing a set of
``valid brackets" from which we then define  the ``generalized Hulth\'en
bracket vectors". They were first introduced by Lieb  and Temperley in
\cite{Temperley1971}. Let $n\in\mathbb N$ be fixed throughout this section. A
valid bracket $\B = ([x_1,y_1],\dots,[x_n,y_n])$ is a collection of $n$
brackets $[x_i,y_i]$ with $1\le x_i<y_i\le L$ for all $i$ and $y_1<\ldots y_n$.
In addition, a valid bracket satisfies the three conditions:
\begin{enumerate}
    \item ({\bf Exclusion}) For each $x \in [1,L]$, let $d_{\B}(x)
    = \#\{i : x_i = x \textrm{ or } y_i=x\}$. Then $d_{\B}(x) \leq
    1$ for all $x \in [1,L]$.
    \item ({\bf Non-crossing}) If $y_i>x_j$ for some $j>i$ then
    $x_i<x_j<y_j<y_i$.
    \item ({\bf Non-spanning}) For any $i$ if there is some $x$
    such that $x_i<x<y_i$, then $d_{\B}(x) = 1$.
\end{enumerate}
We define $|\B|=n$. The set of all valid brackets on the chain $[1,L]$ is 
denoted by $\mathcal{V}([1,L],n)$. Now, given a valid bracket 
$\B\in\mathcal{V}([1,L],n)$, we define the ``generalized Hulth\'en bracket
vector" $\psi(\B) \in \Hil_{[1,L]}$ as
\begin{equation}\label{def:Hbb}
    \psi(\B)\, :=\, \prod_{i=1}^{|\B|} \left(q^{-1/2}\, S_{x_i}^- -
    q^{1/2}\,
    S_{y_i}^-\right) \ket{\Uparrow}_{[1,L]}\, .
\end{equation}
$\ket{\Uparrow}_{[1,L]}$ is, of course, the all up-spin vector in 
$\Hil_{[1,L]} = (\C^2)^{\otimes L}$. Lieb and Temperley proved that for 
any $n \in [0,\floor{L/2}]$, the set $\mathcal{V}([1,L],n)$ forms a basis 
spanning $\Hil^{\HW}([1,L],n)$. Therefore, to calculate $E(L,n)$, it
suffices to find the minimum eigenvalue for the matrix of $H^{\rm
k}_{[1,L]}$ in the basis of $\psi(\B)$. In addition, the action of
$h^{\rm k}_{x,x+1}$ upon these basis vectors is simple and has a very 
appealing graphical representation which we will see after the following
Lemma. 
\begin{lemma}
\label{lem:TL} For $x \in [1,L-1]$ and $\B\in\mathcal{V}([1,L],n)$, let
\begin{equation*}
    \phi\, =\, -[2]_q\, h^{\rm k}_{x,x+1}\, \psi(\B)\, .
\end{equation*}
Then $\phi$ has the following values, depending on the case.
\begin{itemize}
    \item If $d_{\B}(x)=d_{\B}(x+1)=0$ then
    $\phi=0$.
    \item If $d_{\B}(x)=1$ but $d_{\B}(x+1)=0$ then there is
    a bracket in $\B$, $[x_i,y_i]$, with $y_i=x$. Then $\phi = \psi(\B')$
    where $\B'$ is defined relative to $\B$ by the replacement $[x_i,y_i] 
    \to [x,x+1]$.
    \item If $d_{\B}(x)=0$ but $d_{\B}(x+1)=1$ then there is
    a bracket in $\B$, $[x_i,y_i]$, with $x_i=x+1$. Then $\phi = \psi(\B')$
    where $\B'$ is defined relative to $\B$ by the replacement $[x_i,y_i] 
    \to [x,x+1]$.
    \item If $d_{\B}(x) = d_{\B}(x+1) = 1$ then one of four
    possibilities occurs.
    \begin{itemize}
    \item There are two brackets in $\B$, $[x_i,y_i]$ and $[x_j,y_j]$, with
    $y_i=x$ and $x_j=x+1$. Then $\phi = \psi(\B')$, where $\B'$ has
    the replacements $\{[x_i,y_i],[x_j,y_j]\} \to
    \{[x,x+1],[x_i,y_j]\}$.
    \item There are two brackets in $\B$, $[x_i,y_i]$ and $[x_j,y_j]$, with
    $y_i=x$ and $y_j=x+1$. Then $\phi = \psi(\B')$,  where $\B'$ has
    the replacements $\{[x_i,y_i],[x_j,y_j]\} \to \{[x,x+1],[x_j,x_i]\}$.
    \item There are two brackets in $\B$, $[x_i,y_i]$ and $[x_j,y_j]$, with
    $x_i=x$ and $x_j=x+1$. Then $\phi = \psi(\B')$,  where $\B'$ has
    the replacements $\{[x_i,y_i],[x_j,y_j]\} \to
    \{[x,x+1],[y_j,y_i]\}$.
    \item The bracket $[x,x+1]$ is in $\B$. Then $\phi = - [2]_q\,
    \psi(\B)$.
    \end{itemize}
\end{itemize}
\end{lemma}
The proof is left to the reader who should notice that the vector
$(q^{-1/2} S_{x}^- - q^{1/2}S_{y}^-) \ket{\Uparrow}_{[1,L]}$ is just a 
multiple of the singlet state between sites $x$ and $y$.

In the graphical representation proposed by Lieb and Temperley we consider
the graph whose vertex set is the ordered chain $[1,L]$. The edges are 
the edges specified by brackets. Namely, there is an edge between $x$ and 
$y$ if $[x,y]\in\B$. Because of rules (1--3) these are precisely the set of 
partial matchings which can all be made above the line passing through 
$[1,L]$, and such that no two edges intersect, nor does any edge span an 
unpaired vertex. We call the edges ``arcs", and moreover since we specify 
that they should be above the line through $[1,L]$, we call them ``upper
arcs". An example for $L=8$ is
\begin{equation*}
    \{[3,4][2,5],[7,8]\}\, \longrightarrow\,
\boxed{\begin{array}{c} \resizebox{6cm}{!}{\includegraphics{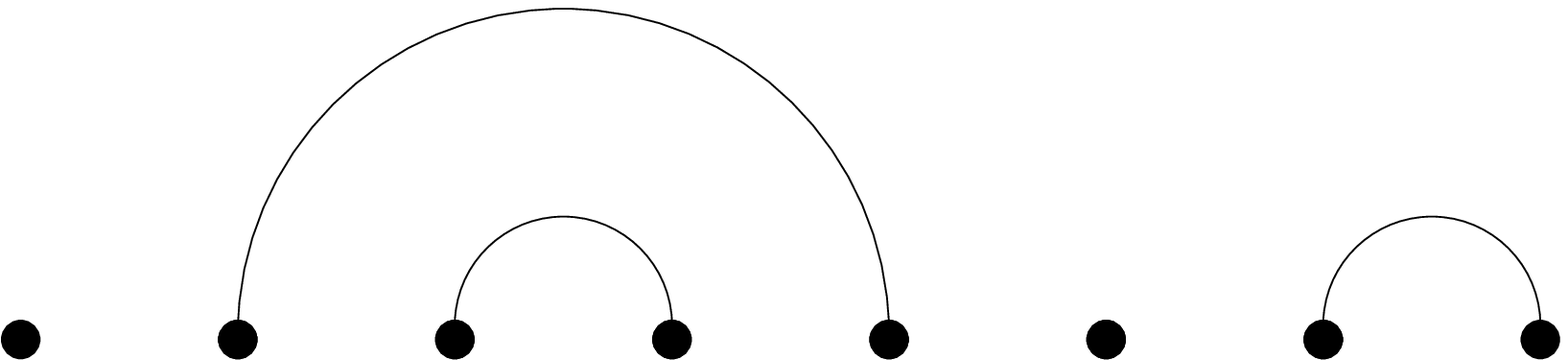}}
\end{array}}
\end{equation*}
Then, if we define $U_{x,x+1} = -[2]_q\, h^{\rm k}_{x,x+1}$, we
have that $U_{x,x+1}$ acts as follows. First of all, associate to
each $U_{x,x+1}$ a diagram of a pair of arcs; e.g,
\begin{align*}
    U_{1,2}\, &=\,
    \boxed{\begin{array}{c} \resizebox{6cm}{!}{\includegraphics{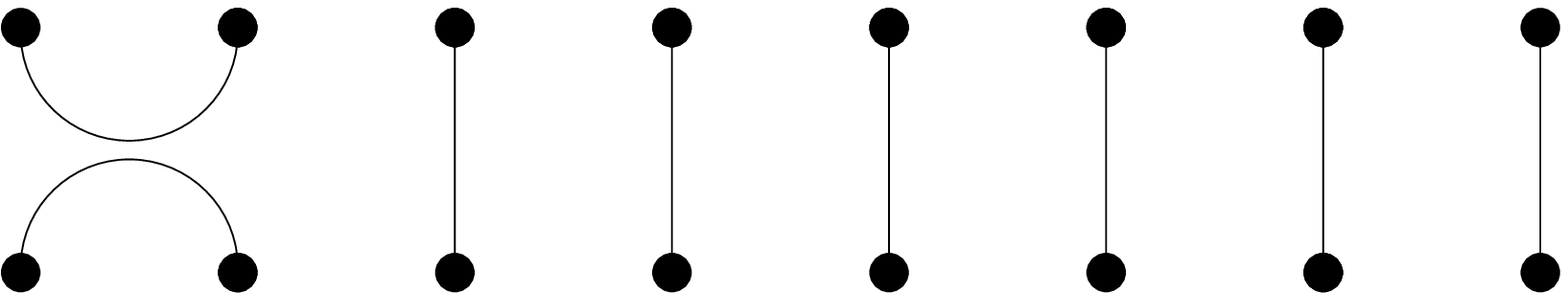}}
    \end{array}}\\
    U_{2,3}\, &=\,
    \boxed{\begin{array}{c} \resizebox{6cm}{!}{\includegraphics{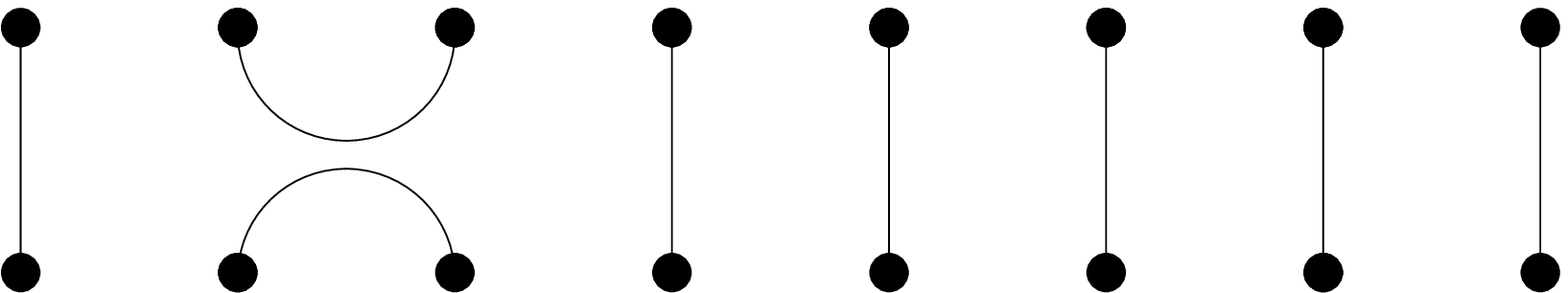}}
    \end{array}}\\
    &\hspace{100pt} \vdots\\
    U_{L-1,L}\, &=\,
    \boxed{\begin{array}{c} \resizebox{6cm}{!}{\includegraphics{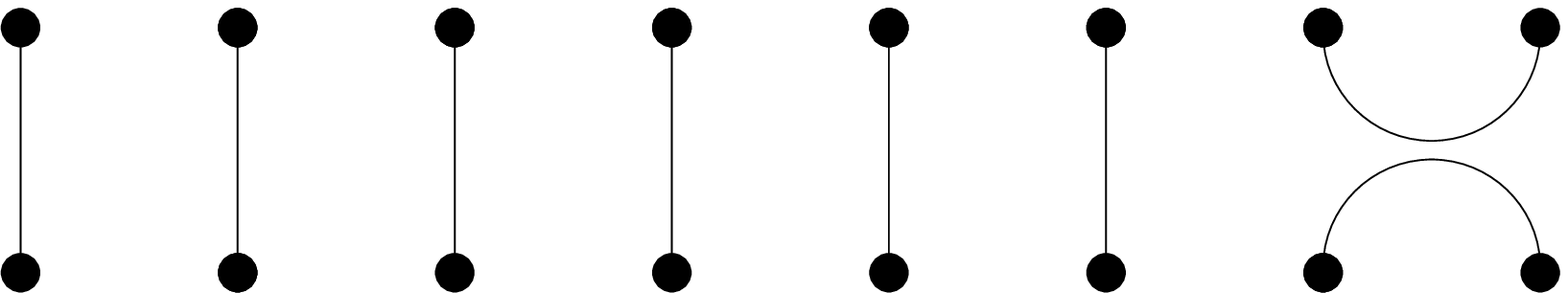}}
    \end{array}}\\
\end{align*}
Imagine concatenating this graph below the arc system for $\B$. Then
contract all loose ends down to vertices, and stretch the arcs to their 
normal shapes. Then one obtains the correct arc system corresponding to 
$\B'$. The one exception is if $\B'=\B$ because $[x,x+1] \in \B$. But in 
this case one obtains $\B$ with one additional circle, or ``bubble". If 
one declares that the bubble takes a scalar value $-[2]_q$ to remove, 
then one has the correct answer for $\phi$ also in this case.
\begin{equation*}
\begin{array}{c} \resizebox{1.5cm}{!}{\includegraphics{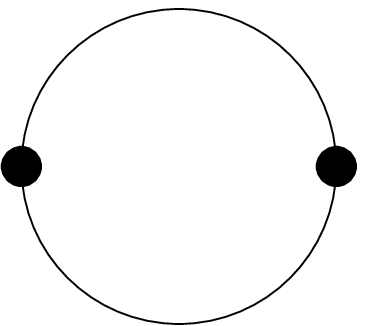}} \end{array}\, 
=\, - (q+q^{-1})\, .
\end{equation*}

At this point let us mention that the matrices $U_{x,x+1} = 
-[2]_q\, h^{\rm k}_{x,x+1}$ satisfy the Temperley-Lieb relations and
that $\Hil_{[1,L](n)}$ is an irreducible representation space of the
Temperley-Lieb algebra. 

\begin{remark}\label{rem:mono}
In this matrix representation, the only dependence on $q$ comes from the
``bubble" terms, which contribute a factor $-[2]_q$ to the diagonal elements.
Since this is clearly decreasing in $q$, it is trivial to see that the minimum
energy of the kink Hamiltonian in any $n$-magnon subspace is decreasing in $q$.
This is pertinent to Remark \ref{rem:cont}.
\end{remark}

\subsection{The kink Hamiltonian on $\Z$ in the Hulth\'en bracket basis}

The purpose of this section is to define the kink Hamiltonian on the
Hulth\'en bracket basis vectors for brackets on $\Z$ instead of $[1,L]$. 
So let us first extend the notion of valid brackets and set 
$\mathcal{V}([-L,L],n)$ to be the set of valid brackets 
$\B=([x_1,y_1],\ldots,[x_n,y_n])$ on $[-L,L]$ satisfying the same 
conditions (1--3) from above. Then we define the set of valid brackets 
on $\Z$, $\mathcal{V}(\Z,n) = \bigcup_{L\in\N}\mathcal{V}([-L,L],n)$. 

For any $\B\in\mathcal{V}(\Z,n)$ we take some $L$ so that $[-L,L]$ 
contains all points in $\B$ and then unambiguously define the
Hulth\'en bracket basis vectors $\psi({\B})$ as in (\ref{def:Hbb})
and tensor with the all spin-up vector on $\Z\setminus[-L,L]$. 

Now, for any $x\in[-L,L-1]$ and $\B\in\mathcal{V}([-L,L],n)$ we define 
the matrix representation,
\begin{equation*}
    h^{\rm k}_{x,x+1}\, \psi(\B)\,
    =\, \sum_{\B' \in \mathcal{V}([-L,L],n)}\,A_{L}^{x}(\B',\B) \,
    \psi(\B')\, .
\end{equation*}
Note that for any such lattice point $x$ and bracket $\B$ we have 
the following fact: If $L'\geq L$, then by Lemma \ref{lem:TL}:
\begin{itemize}
    \item If $\B' \in \mathcal{V}([-L,L],n)$ then
    $A_{L'}^x(\B',\B) = A_{L}^x(\B',\B)$.
    \item If $\B' \in \mathcal{V}([-L',L'],n) \setminus
    \mathcal{V}([-L,L],n)$ then $A_{L'}^x(\B',\B)=0$.
\end{itemize}
Thus we can define for any $x\in\Z$ the kernel $A^x$ by
\begin{equation*}
    A^x(\B',\B)\, =\, \lim_{L \to \infty}
    A_{L}^x(\B',\B)\, .
\end{equation*}
Moreover, by the same rules explained in Lemma \ref{lem:TL}, we have
\begin{itemize}
    \item If $\B \in \mathcal{V}([-L,L],n)$ and $x \in \Z \setminus
    [-L-1,L]$, then $A^x(\B',\B) = 0$ for every $\B' \in \mathcal{V}(\Z,n)$;
    \item If $\B' \in \mathcal{V}([-L,L],n)$ and $x \in \Z
    \setminus [-L,L]$, then $A^x(\B',\B) = 0$ for every $\B \in
    \mathcal{V}(\Z,n)$.
\end{itemize}
Therefore, we can define a valid, finite kernel, $A :
\mathcal{V}(\Z,n) \times \mathcal{V}(\Z,n) \to \R$ by
\begin{equation*}
    A(\B',\B)\, =\, \sum_{x \in \Z} A^x(\B',\B)\, ,
\end{equation*}
since all but a finite number of summands will be zero. Moreover,
from this we conclude that if $\B \in \mathcal{V}([-L,L],n)$, then
$A(\B',\B)=0$ unless $\B' \in \mathcal{V}([-L-1,L],n)$. Therefore,
defining $\ell(\mathcal{V}(\Z,n))$ to be the set of all sequences
on the countable set $\mathcal{V}(\Z,n)$, we may define the linear
transformations ${\mathcal A}_n$ and the infinite chain kink Hamiltonian 
$H_{\Z,n}^{\rm k}$ by
\begin{eqnarray*}
    {\mathcal A}_n\delta_\B\, &=&\, \sum_{\B' \in \mathcal{V}(\Z,n)} 
    A(\B',\B)\,\delta_{\B'}\, ,
    \\
    H_{\Z,n}^{\rm k}\psi(\B)\, &=&\, \sum_{\B' \in \mathcal{V}(\Z,n)} 
    A(\B',\B)\,\psi({\B'})\, .
\end{eqnarray*}
Again, all but finitely many terms in the sum are zero. Since 
${\mathcal A}_n$ and $H_{\Z,n}^{\rm k}$ are defined by the same kernel 
they have the same spectrum. 

\begin{lemma} \label{lem:Abounded}
The linear operator ${\mathcal A}_n$ restricted to $\ell^2(\mathcal{V}(\Z,n))$
is bounded.
\end{lemma}

\begin{proof} For this we will use again the Riesz-convexity 
theorem. From Lemma \ref{lem:TL}, we observe the following facts:
\begin{itemize}
    \item If $\B = ([x_1,y_1],\dots,[x_n,y_n])\in \mathcal{V}(\Z,n)$, 
    then $A^x(\B',\B)$ is zero unless $|x-x_i|\leq 1$ or $|x-y_i|\leq 1$ 
    for some $i \in [1,n]$. In any of these cases, there is exactly one 
    $\B'$ such that $A^x(\B,\B')$ is nonzero.
    \item If $\B'= ([x_1',y_1'],\dots,[x_n',y_n']) \in \mathcal{V}(\Z,n)$, 
    then $A^x(\B',\B)$ is
    zero unless $x=x_i'$ and $x+1=y_i'$ for some $i \in [1,n]$. In this case,
    the only way for $A^x(\B',\B)$ to be nonzero is if either: $\B=\B'$; or
    if $\B$ differs from $\B'$ by the replacement of the arc
    $[x_i',y_i']$ by another arc with one endpoint in $\{x,x+1\}$;
    the replacement of the arc $[x_i',y_i']$ and another arc,
    bracketing it, $[x_j',y_j']$ by the arcs $[x_j',x_i']$ and
    $[y_i',y_j']$. The total number of these possibilities is bounded
    by $n$.
    \item Whenever $A^x(\B,\B')$ is nonzero, the absolute value is bounded 
    by $1$. 
\end{itemize}
Using these facts, one can conclude that $\mathcal{A}_n$ is bounded both 
in $\ell^1(\mathcal{V}(\Z,n))$ and $\ell^\infty(\mathcal{V}(\Z,n))$
by bounding the maximum ``row sum" and ``column sum" of $A$.
Therefore, by the Riesz-convexity theorem, it happens that $\mathcal{A}_n$ 
is bounded on every $\ell^p(\mathcal{V}(\Z,n))$ for $1\leq p\leq
\infty$. In particular, it works for $p=2$.
\end{proof}

\subsection{Bounds for the change-of-basis transformation}
\label{sec:R}

We show now that the change from the Hulth\'en bracket basis to the Ising
basis in the $n$-magnon sector $\Hil_\Z(n)$ is bijective and bounded.
It is important that we do this on $\Z$ and to remember that $q<1$.
We use the notation from Section \ref{Section 3}. 

\begin{lemma} \label{lem:transf}
Define a transformation $R : \ell^p(\mathcal{V}(\Z,n)) \to 
\ell^p(\mathcal{X}_n)$ such that
\begin{equation*}
    \psi(\B)\, =\, \sum_{\boldsymbol{x}\in\mathcal{X}_n} 
    \d_{\boldsymbol{x}} \,R(\B,\boldsymbol{x})\,.
\end{equation*}
Then $R$ is invertible and bounded for $p=1$ and $p=\infty$. More precisely,
$$ \| R \|_{1} \,\le\, q^{-1/2} \frac{(2n)!}{n!}\,,
   \quad \mbox{and}\quad
   \| R \|_{\infty} \,=\, \left(q^{-1/2} + q^{1/2}\right)^n \,.
$$
Moreover, $R$ intertwines between the operator $\mathcal{A}_n$ 
and $H_{\Z}(n)$, namely, $R \mathcal{A}_n = H_\Z(n) R$.
\end{lemma}

\begin{proof} The intertwining property follows immediately since the  
kink boundary fields will telescope to 0 in the $L \to \infty$ limit of 
the all-up-spin GNS representation.

The colums of $R$ represent the coefficients
of $\psi(\B)$ in the expansion in the Ising basis. The map $R$ is 
clearly injective. On the other hand, given an Ising vector 
$\d_{\boldsymbol{x}}$, we can write this as a norm-convergent
telescoping sum of Hulth\'en bracket vectors. This is not possible 
on a finite chain nor when $q=1$. 

Recall the following basic fact. If $B:\C^{n_1}\to
\C^{n_2}$ is a linear transformation, and 
$$ (Bu)_i = \sum_{j=1}^{n_1} B_{ij} u_j\,,\quad i=1,\ldots, n_2\,. 
$$
Then 
$$ \|B\|_{\infty} = \max_{1\le i \le n_2} \sum_{j=1}^{n_1} 
   |B_{ij}|\,.
$$
Furthermore, $\|B\|_1\le\|B^t\|_\infty$. [Although $R$ itself is not a 
finite dimensional matrix, we could restrict the discussion of the 
change-of-basis transformation to finite chains where we would
show that our bounds are uniform in $L$.]  

If we expand $\psi(\B)$ in the Ising basis $\d_{\boldsymbol{x}}$, then
we see that the sum of the absolute values of the coefficients is
$\left(q^{-1/2} + q^{1/2}\right)^n$, and we therefore have that
$\| R \|_{\infty} = \left(q^{-1/2} + q^{1/2}\right)^n$. 

The other bound is combinatorial and rests on the following claim.

\medskip
{CLAIM:} {\em Let $\boldsymbol{w} = (w_1,\dots,w_n)$ be a collection of 
points in $\mathcal{X}_n$. There is an upper bound on the number of 
nonspanning systems of arcs
$$
\B = ([x_1,y_1],\ldots,[x_n,y_n])\, ,
$$ 
for which there exists a permutation $\pi \in \mathcal{S}_n$ such that 
$w_{\pi(k)} \in \{x_k,y_k\}$ for each $k\in[1,n]$. The bound is $(2n)!/n!$.}

\medskip

Given this estimate we instantly get 
$$ \|R\|_1 \le \|R^t\|_\infty \le q^{-1/2} \frac{(2n)!}{n!}\,.
$$

\smallskip
{Proof of CLAIM:}
Suppose that $\B = ([x_1,y_1],\dots,[x_n,y_n])$ is a 
nonspanning system of arcs (i.e., satisfying conditions 1 and 3) but
possibly with crossings, and such that there exists a $\pi \in 
\mathcal{S}_n$, such that $w_{k} \in \{x_{\pi(k)},y_{\pi(k)}\}$
for all $k \in [1,n]$. Let $\tilde{v}_k$ be the complementary point so that 
$\{w_{k},\tilde{v}_k\} = \{x_{\pi(k)},y_{\pi(k)}\}$ for each $k$.
Let $(v_1,\dots,v_n)$ be the rearrangement of $\tilde{v}_1,\dots,\tilde{v}_n$
in increasing order. Then we claim that the arc system $\tilde{\B}$,
whose arcs are (the ordered rearrangement of) $\{w_{k},v_{k}\}$ is also 
nonspanning. The reason is that one can transpose the $(\tilde{v}_k,
\tilde{v}_{k+1})$ such that $\tilde{v}_k>\tilde{v}_{k+1}$, one-at-a-time 
to obtain the desired goal, and never span new sites.
Particularly, the only new sites which could be spanned by transposing
$\tilde{v}_k > \tilde{v}_{k+1}$ are the sites in
$[\tilde{v}_k,\tilde{v}_{k+1}]$. But since $w_k<w_{k+1}$, these sites 
must have already been spanned by one (or both) of the two arcs whose 
endpoints are (the ordered rearrangements of)
$\{\tilde{v}_k,w_k\}$ and $\{\tilde{v}_{k+1},w_{k+1}\}$.

Now, we can determine a set of points $\{u^{-}_1,\dots,u^{-}_n,u^{+}_1,
\dots,u^{+}_n\}$ such that $\{v_1,\dots,v_n\}$ must be a subset of this one.
This would obviously prove the claim, since there are at most $(2n)!/(n!)^2$
such subsets, and $\B$ is uniquely determined by $(\tilde{v}_1,\dots,
\tilde{v}_n)$ which is obtained from $(v_1,\dots,v_n)$ by permuting by one 
of the $n!$ permutations in $\mathcal{S}_n$.

Let $\{u_1^{-},\dots,u_n^{-}\}$ be the points such that
$u_n^{-}$ is the first point to the left of $w_n$ which is not 
among $\{w_1,\dots,w_n\}$, and for each $k<n$, the point $u^{-}_k$ is 
the first point to the left of $w_k$ which is not among 
$\{w_1,\dots,w_n\}$ or $\{u^{-}_{k+1},\dots,u^{-}_{n}\}$.
We claim that if $v_k<w_k$ then it must be among
$\{u^{-}_k,\dots,u^{-}_n\}$. This is because, in this case,
$\{v_k,w_k\}$ spans no sites other than $\{w_1,\dots,w_n\}$
or $\{v_1,\dots,v_n\}$. Since $v_j<v_k$ for $j<k$, in fact it spans 
no sites other than $\{w_1,\dots,w_n\}$ and $\{v_k,\dots,v_n\}$.
A similar construction for $\{u^{+}_1,\dots,u^{+}_n\}$
allows the conclusion that if $v_k>w_k$ then
$v_k$ is in the set $\{u^{+}_1,\dots,u^{+}_k\}$.

\end{proof}

\remarks \begin{enumerate}
\item By the open mapping theorem, $R^{-1}$ is also bounded. 
\item From the Riesz-convexity theorem we derive the bound
$$ 
\|R\|_p \le \|R\|_1^{1/p} \, \|R\|_\infty^{1/q} 
$$
for any $p\in[1,\infty]$ with $1/p+1/q =1$. In particular, the map 
$$
R:\ell^2(\mathcal{V}(\Z,n))\to\Hil_\Z(n)
$$ 
and its inverse are bounded.
\end{enumerate}

\subsection{A Wielandt theorem}

This section extends a Wielandt-type theorem~\cite{Nachtergaele1987} 
applicable to Banach spaces. We actually prove a stronger statement then
needed. So let us consider a countable set
${X}$. Then, let $k:{X}\times{X} \to \R$ be a 
kernel with the following properties
\begin{enumerate}
\item There exists a uniform $k_0<\infty$ so that $0\leq k(x,y) \leq k_0$ 
for all $x,y\in{X}$;
\item There is an integer $N$ such that
\begin{equation*}
   \sup_{y \in {X}} \#\{x \in {X}\, :\, k(x,y)\neq
   0\}\, \leq N\, ,\mbox{and}
   \sup_{x \in {X}} \#\{y \in {X}\, :\, k(x,y)\neq
   0\}\, \leq N\, .
\end{equation*}
\end{enumerate}
Interpolating between $\ell^1$ and $\ell^\infty$, we know from the 
Riesz-convexity theorem that this kernel defines a linear bounded 
operator, ${K}:\ell^2({X})\to\ell^2({X})$. The first partial result
concerns the spectral radius of a restriction of $K$. So let
${Y}\subset{X}$. Then we define the operator ${K}\!\restriction\! 
{Y}$ to be the operator on $\ell^2({Y})$ whose kernel is
$k\!\restriction\! {Y}\times{Y}$. 

\begin{prop}[Generalized Wielandt theorem] \label{lem:wie}
We assume the same
conditions on $K$ as above. Let $Y$ be a finite subset of $X$ and let
$j\geq0$ be a kernel on $Y\times Y$. Let $J$ be the operator on $\ell(Y)$
defined by this kernel. If $j(x,y) \leq k(x,y)$ for all $(x,y) \in 
Y \times Y$, then $\rho(J) \leq \rho(K)$.
\end{prop}

\begin{remark}
The norm for $\ell(Y)$ is immaterial since $Y$ is a finite set.
\end{remark}

\begin{proof} 

By the standard Perron-Frobenius theorem for matrices, there is a vector 
$\psi \in \ell(Y)$ with eigenvalue $\lambda=\rho(Y)$.

By extending it to be zero on $X \setminus Y$, we can also consider
this as a vector in $\ell^2(X)$. Moreover, by the our hypotheses, 
we have that $J\psi \leq K\psi$. This implies that $(K-\lambda)\psi\geq0$.
Writing $(K^{n}-\lambda^{n})\psi = K(K^{n-1}-\lambda^{n-1})\psi + 
\lambda^{n-1} (K-\lambda)\psi$ we conclude inductively that $K^n\psi\ge
\lambda^n\psi$ for all $n\in\N$. Since the kernel has positive entries, we
get that $\|K^n\psi\| \geq \lambda^n \|\psi\|$. Therefore, $\|K^n\|\geq
\lambda^n$, and
$$
  \rho(K)\, =\, \lim_{n \to \infty} \|K^n\|^{1/n}
  \, \geq\, \lambda\, =\, \rho(J)\, .
$$
\end{proof}

\begin{remark}
This proposition and proof follow \cite{Nachtergaele1987}.
\end{remark}

\subsection{Proof of the main proposition}
\label{Sect:proof of kinkgs}

The last item of our business is to prove Proposition 
\ref{prop: kinkgs} about the ground state energies of the kink 
Hamiltonian. Since $H_n$ is self-adjoint, it follows that
\begin{equation*}
    \infspec(H_{\Z,n})\, =\, \inf_{\substack{\psi \in \Hil_\Z(n)\\
    \|\psi\|=1}} \ip{\psi}{H_{\Z,n} \psi}\, .
\end{equation*}
On the other hand, using Lemma \ref{lem:transf}, we can write
\begin{equation*}
    \infspec(H_{\Z,n})\, =\, \inf_{\substack{\phi \in 
    \ell^2(\mathcal{V}(\Z,n))\\
    \|R \phi\|=1}} \ip{\phi}{\mathcal{A}_n \phi}\, .
\end{equation*}
Moreover, with the natural identification of
$\mathcal{V}([-L,L],n) \subset \mathcal{V}(\Z,n)$, we have density
\begin{equation*}
    \ell^2(\mathcal{V}(\Z,n))\,
    =\, \operatorname{cl}\left(\bigcup_{L \in \N}
    \ell^2(\mathcal{V}([-L,L],n))\right)\, .
\end{equation*}
Therefore,
\begin{equation*}
    \infspec(H_{\Z,n})\, =\, \inf_{L \in \N}\, \min_{\substack{\phi \in
    \ell^2(\mathcal{V}([-L,L],n))\\ \|R\phi\|=1}}
    \ip{\phi}{\mathcal{A}_n \phi}\, .
\end{equation*}
Now, let $\varepsilon>0$. Since both $H_{\Z,n}$ and $\mathcal{A}_n$ are 
bounded below, there does exist an $L \in \N$ and $\phi \in
\ell^2(\mathcal{V}([-L,L],n))$ such that $\|R\phi\|=1$ and
$\ip{\phi}{\mathcal{A}_n \phi} \leq \infspec(\mathcal{A}_n) + \epsilon$. 
For such a vector $\phi \in\ell^2(\mathcal{V}([-L,L],n))$ we have a unique
vector $\psi\in\Hil^{\HW}([-L,L],n)$ with the property that $\|\psi\|=1$ and
$\ip{\phi}{\mathcal{A}_n \phi} = \big(\psi,{H_{[-L,L]}^{\rm k} \psi}\big)$. By 
shifting the interval $[-L,L]$ to the right by $L+1$ units, we conclude that
\begin{equation*}
    E(2L+1,n)\, \leq\, \varepsilon + \infspec(H_{\Z,n})\, .
\end{equation*}
Since $\varepsilon$ was arbitrary, and using the monotonicity of
$E(L,n)$ in $L$, we have
\begin{equation*}
    E(n)\, \leq\, \infspec(H_{\Z,n})\, .
\end{equation*}

For the opposite inequality we use Proposition \ref{lem:wie}. Let 
$X=\mathcal{V}(\Z,n)$. Let $L\in \N$, and ${Y} = \mathcal{V}([-L,L],n)$.
Then we consider the matrix $B:{Y}\times{Y} \to \R$ given by the kernel
\begin{equation*}
    B(\B',\B)\, =\, \sum_{x=-L}^{L} A_{L}^x(\B',\B)\, .
\end{equation*}
The operators $J=n-B$ and ${K} = n - \mathcal{A}_n$ satisfy the 
conditions of  Proposition \ref{lem:wie}. Therefore, we conclude that
\begin{equation*}
    \rho(n-B)\, \leq\, \rho(n-\mathcal{A}_n)\, .
\end{equation*}
But using the fact that the spectra of $B$ and ${K}$ are both real
subsets (because the associated operators are similar to
self-adjoint operators), we conclude that
\begin{equation*}
    \rho(n-B)\, =\, n - \infspec(B)\quad \textrm{and}\quad
    \rho(n-{K})\, =\, n - \infspec({K})\, .
\end{equation*}
Therefore,
\begin{equation*}
    E(L,n)\, \geq\, \infspec(A_n)\, ,
\end{equation*}
as desired.

\appendix
\section{Results for small $q$}

In this appendix, we collect some results for small $q$.  The primary purpose
of this is to verify the Bethe ansatz formulas for droplet eigenstates of the
reduced Hamiltonian $H(n,\theta)$ for other values of $\theta \in \mathbb{S}^1$
than $\theta = 0$. Using the methods of Section \ref{sec:dropletcyclic} we can
also treat the cyclic Hamiltonian $H^{{\rm cyc}}$ in the $e^{i\theta}$
eigenspaces of the translation operator. In particular, the latter is
interesting because this is the framework analyzed by Yang and Yang in
\cite{Yang1966c}.   Previously, this regime of the XXZ model (small $q$ and
cyclic boundary conditions) was rigorously analyzed by Kennnedy in
\cite{Kennedy2005}, using the methods developed in \cite{Kennedy2003}. (The
purpose of \cite{Kennedy2005} was partly to give a pedagogic introduction to
the methods of \cite{Kennedy2003}, but it also gave new and interesting results
for the XXZ model, some of which we describe below.)

Before going further, we would like to mention that in the paper proper, none
of the arguments were perturbative.  All applied to the entire region $q \in
(0,1)$, which is the maximal interval where the results are valid. This is
important to keep in mind when one considers the relatively simple arguments to
follow.

\subsection{Droplet energies in the infinite chain for small $q$}

Let us fix $n\in\N_+$. Before stating the main result of this section, we
recall the following. The Hamiltonians $H_{\Z}(n,\theta)$ are periodic of
period $2\pi/n$ in the sense that there is a unitary phase multiplication, as
in (\ref{constraint}), such that after conjugating by that $H_{\Z}(n,\theta)$
and $H(n,\theta+2\pi/n)$ are equal. In particular, this means that the spectrum
is $2\pi/n$ periodic. Moreover, if there is an eigenvector of
$H_{\Z}(n,\theta)$, then multiplying this eigenvector by the necessary phase
produces the relevant eigenvector for $H(n,\theta+2\pi/n)$.

\begin{prop}
\label{prop:smallq_infinite}
There exists a constant $q_0=q_0(n)>0$ such that for $0<q<q_0$, the infspec of
$H_{\Z}(n,\theta)$ is an eigenvalue for all $\theta$, and for $\theta \in
(-\pi/N,\pi/N)$ the eigenvector is the one given in Lemma \ref{lem:Bethe}. The
eigenvectors are norm continuous in $\theta$, and are determined for all
$\theta$ using this and periodicity. Moreover, there is a constant
$\gamma(n,q)>0$ such that there  is a spectral gap above the ground state of
$H_{\Z}(n,\theta)$ of size at least $\gamma(n,q)$, uniformly in $\theta$. 
\end{prop}

\begin{proof}
Fixing $n$ and $\theta$, there is obviously a spectral gap above the bound
state for the Ising model, $q=0$. It is easy to see that the gap is 1 at $q=0$.
But the kernel $K_{n,\theta}$, when thought of as a function of $q$, varies in
a way such that the associated operators are norm-continuous with respect to
$q$, on $\ell_0^2(\mathcal{X}_n)$. (As used before in the paper, this can be
proved by obtaining row and column sum bounds, which pertain to $\ell^1$ and
$\ell^\infty$, and then using Riesz convexity.) Therefore, there is some $q_0$
and some curve $\gamma(n,q)$, positive for $q<q_0$, such that
$H_{\Z}(n,\theta)$ has a unique ground state and a spectral gap of size at
least $\gamma(n,q)$ for all $\theta$ as long as $0<q<q_0$. But the bound states
found in Lemma \ref{lem:Bethe} vary continuously in $q$, therefore, they must
be the actual eigenstates. \end{proof}

\begin{remark}
The argument of the proof is, to some extent, an analogue of Yang and Yang's
argument from \cite{Yang1966a} but starting from the Ising model, not the XY
model, and valid directly in the infinite volume limit. We would like to
mention that more sophisticated and more powerful arguments of the Yang, Yang
style were employed by Goldbaum in \cite{Goldbaum2005} to handle the more
complex -- but still Bethe ansatz solvable --Hubbard model.
\end{remark}

\begin{remark} 
Note that in the proposition above, one cannot choose $q_0$ to
be independent of $n$. The reason is that in our $\ell^1$, $\ell^\infty$
interpolation, the rowsums and columnsums do depend on $n$ because of the {\em
number of} matrix entries. On the other hand, in \cite{Nachtergaele2001} two of
the authors proved a positive spectral gap for all $q$ and $n$ with $q^n$ small
enough, which is uniform in this regime. Therefore, using that result and the
present argument, one can obtain a single $q_0$ which works for all $n>0$.
\end{remark}

Note that not only are the energies for the bound states continuous in $\theta$
and $q$, they are easily seen to be analytic. This is simply because the kernel
entries of the operator are analytic in $\theta$ and $q$, and using the
properties of the kernel (that there are a finite number of nonzero entries in
each row and column) we deduce analyticity of ($q$-dependent) Fourier-reduced
Hamiltonian $H_q(n,\theta)$ in the weak-topology. Using the spectral gap this
is sufficient to guarantee analyticity of the eigenvectors.  Using analyticity
in $\theta$, we can obtain the following result.

\begin{cl}  
The spectrum of $H_{\Z}(n)$ in the range $(0,\gamma_{l}(n,q))$ is
absolutely continuous. 
\end{cl}

We will not give a detailed proof, but the reader is referred to Theorem
XIII.86 of \cite{Reed1978}. After conjugating by the spectral projection onto
$(0,\gamma(n,q))$ the Hamiltonian satisfies the conditions of that theorem.

\begin{remark}  
One probably expects that the entire spectrum of $H_{\Z}(n)$ is
absolutely continuous for all $0<q\leq 1$. Using the results of
\cite{Babbitt1977b} this is presumably provable at $q=1$. But in general the
translation-invariance suggests it is true. 
\end{remark}

\subsection{The Hamiltonian for the cyclic chain for small $q$}

Note that for the finite cyclic chain, just as for the infinite chain, there is
a well-defined translation operator, commuting with the Hamiltonian. The
following result was proved by Kennedy in \cite{Kennedy2003}.

\begin{prop}  
There exists a $q_0$ such that for $0\leq q\leq q_0$ the $L$
lowest energy levels of $H_{[1,L]}^{\textrm{cyc}}$ in the sector with $n$
downspins ($0<n<L$) can be indexed by the translation eigenvalues
$e^{i\theta}$, for $\theta=2\pi k/L$ and $k \in \Z/L\Z$. For all $\theta$,
there is an analytic expression for the energy eigenvalue
$E^{\textrm{cyc}}(L,n,\theta)$ satisfying \begin{equation*} \lim_{L \to \infty}
E^{\textrm{cyc}}(L,n,\theta)\, =\, 1 + \sum_{s=-\infty}^{\infty} d_s e^{i\theta
s}\, . \end{equation*} The coefficients $d_s=d_s(n,q)$ are of order $O(q^n)$.
\end{prop}

\begin{remark}
The arguments in \cite{Nachtergaele2001} prove that there is also a gap for
large enough $n$ and small enough $q$, and calculates the asymptotic form of
the energy in the $n \to \infty$ limit, with $q$ fixed. As is easily seen from
our present analysis, in that limit the energy converges to $\alpha =
(1-q^2)/(1+q^2)$. A simpler argument, but which is not robust to changes in
$L$, can follow the proof of Proposition \ref{prop:smallq_infinite}. Namely,
one can construct a kernel in each subspace of $n$ downspins and translation
eigenvalue $\theta$, and check that as a function of $q$ the kernel is
continuous, and moreover it is uniformly continuous for $q \in [0,1]$. If one
considers the sequence of operators for all $L$ (acting on different Hilbert
spaces depending on $L$) one can even deduce that they are in some sense
equicontinuous, because the stronger fact is true that the operators are
Lipschitz with Lipschitz constants which are uniformly bounded in $L \in \N_+$
and $q \in [0,1]$.
\end{remark}

\begin{remark}
The important technique of Kennedy, which follows the previous work
\cite{Kennedy2003}, is to obtain a perturbation expansion which can be
performed for all $L$ at once, therefore allowing comparison of different $L$. 
\end{remark}

\begin{cl}
The series expansion of Kennedy for the $L\to\infty$ limit mat\-ches the 
analytic expressions obtainable from Lemma \ref{lem:Bethe}.
\end{cl}
	
\begin{proof}
One wants to show that
\begin{equation*}
\lim_{\substack{L \to \infty \\ \theta_n \to \theta}} 
E^{\textrm{cyc}}(L,n,\theta_L)\,
=\, E_{\Z,n}(\theta)\, .
\end{equation*}
One knows the existence of  a spectral gap in the $\theta_n$ subspaces for 
small enough $q$ uniform in $L$ for $H^{\textrm{cyc}}_{[1,L]}$. In the last 
subsection, we established a similar spectral gap for $H_{\Z}(n,\theta)$ on 
the infinite chain. Therefore, we can use exactly the same argument as in 
Section \ref{sec:dropletcyclic} to establish the same result for all 
$\theta  \in \mathbb{S}^1$ that we established for $\theta=0$, there: namely 
Proposition  \ref{prop:cycl}. The reader will find that 
translation invariance played no special r\^ole in that argument.
\end{proof}

\par\noindent
{\em Acknowledgement.} B.N. acknowledges  support and hospitality
from the Erwin Schr\"odinger Institute for Mathematical Physics, Vienna, 
and the Centre de Phy\-sique Th\'eorique,
Luminy, where part of this work was carried out.

\bibliographystyle{plain}

\begin{thebibliography}{10}

\bibitem{Babbitt1990}
D.~Babbitt and E.~Gutkin.
\newblock The {P}lancherel formula for the infinite {XXZ} {H}eisenberg spin
  chain.
\newblock {\em Lett. Math. Phys.}, 20(2):91--99, 1990.

\bibitem{Babbitt1977b}
D.~Babbitt and L.~Thomas.
\newblock Ground state representation of the infinite one-dimensional
  {H}eisenberg ferromagnet. {II:} {A}n explicit {P}lancherel formula.
\newblock {\em Commun. Math. Phys.}, 54(3):255--278, 1977.

\bibitem{Faddeev1981}
L.D. Faddeev and L.A. Takhtajan.
\newblock What is the spin of a spin wave?
\newblock {\em Phys. Lett.}, 85A:375--377, 1981.

\bibitem{Goldbaum2005}
P.~Goldbaum.
\newblock Existence of {Solutions} to the {Bethe Ansatz Equations} for the {1D
  Hubbard Model: Finite Lattice and Thermodynamic Limit}.
\newblock {\em Commun. Math. Phys.}, 258(2):317--337, 2005.

\bibitem{Kassel1995}
C.~Kassel.
\newblock {\em Quantum Groups}.
\newblock Springer Verlag, 1995.

\bibitem{Kennedy2005}
T.~Kennedy.
\newblock Expansions for {D}roplet {S}tates in the {F}erromagnetic {XXZ}
  {H}eisenberg {C}hain.
\newblock {\em Markov Processes and Rel. Fields}, 11:223--236, 2005.

\bibitem{Kennedy2003}
T.~Kennedy and N.~Datta.
\newblock Instability of interfaces in the antiferromagnetic {XXZ} chain at
  zero temperature.
\newblock {\em Commun. Math. Phys.}, 236(3):477--511, 2003.

\bibitem{Koma1997}
T.~Koma and B.~Nachtergaele.
\newblock The spectral gap of the ferromagnetic {XXZ} chain.
\newblock {\em Lett. Math. Phys.}, 40:1--16, 1997.

\bibitem{Koma1998}
T.~Koma and B.~Nachtergaele.
\newblock The complete set of ground states of the ferromagnetic {XXZ} chains.
\newblock {\em Adv. Theor. Math. Phys.}, 2:533--558, 1998.

\bibitem{Lu1993}
S.L. Lu and H.T. Yau.
\newblock Spectral gap and logarithmic {Sobolev} inequality for {Kawasaki} and
  {Glauber} dynamics.
\newblock {\em Commun. Math. Phys.}, 156:399--433, 1993.

\bibitem{Matsui1996}
T.~Matsui.
\newblock On ground states of the one-dimensional ferromagnetic {XXZ} model.
\newblock {\em Lett. Math. Phys.}, 37:397, 1996.

\bibitem{Nachtergaele1996}
B.~Nachtergaele.
\newblock The spectral gap for some spin chains with discrete symmetry
  breaking.
\newblock {\em Commun. Math. Phys.}, 175:565--606, 1996.

\bibitem{Nachtergaele1987}
B.~Nachtergaele and L.~Slegers.
\newblock Construction of equilibrium states for one-dimensional classical
  lattice systems.
\newblock {\em Il Nuovo Cimento}, 100 B:757--778, 1987.

\bibitem{Nachtergaele2004}
B.~Nachtergaele, W.~Spitzer, and S.~Starr.
\newblock Ferromagnetic ordering of energy levels.
\newblock {\em Journ. Stat. Phys.}, 116:719--738, 2004.

\bibitem{Nachtergaele2001}
B.~Nachtergaele and S.~Starr.
\newblock Droplet states in the {XXZ} {Heisenberg} model.
\newblock {\em Commun. Math. Phys.}, 218:569--607, 2001.

\bibitem{Reed1978}
M.~Reed and B.~Simon.
\newblock {\em Methods of Modern Mathematical Physics, vol 4. Analysis of
  Operators}.
\newblock Academic Press, San Diego, CA, 1978.

\bibitem{Spitzer2003}
W.~Spitzer and S.~Starr.
\newblock Improved bounds on the spectral gap above frustration free ground
  states of quantum spin chains.
\newblock {\em Lett. Math. Phys.}, 63:165--177, 2003.

\bibitem{Stein}
E.~M.\ Stein and G.~Weiss.
\newblock {\em Introduction to Fourier Analysis on Euclidean Spaces}.
\newblock Princeton University Press, Princeton, NJ, 1971.

\bibitem{Temperley1971}
H.~N.~V. Temperley and E.~H. Lieb.
\newblock Relations between the `percolation' and `colouring' problem and other
  graph-theoretical problems associated with regular planar lattices: some
  exact results for the `percolation' problem.
\newblock {\em Proc.\ Roy.\ Soc.}, A322:252--280, 1971.

\bibitem{Yang1966a}
C.N. Yang and C.P. Yang.
\newblock One-dimensional {C}hain of {A}nisotropic {S}pin-{S}pin
  {I}nteractions. {I}. {P}roof of {B}ethe's {H}ypothesis for {G}round {S}tate
  in a {F}inite {S}ystem.
\newblock {\em Phys. Rev}, 150(1):321--327, 1966.

\bibitem{Yang1966c}
C.N. Yang and C.P. Yang.
\newblock One-dimensional {C}hain of {A}nisotropic {S}pin-{S}pin
  {I}nteractions. {III}. {Applications}.
\newblock {\em Phys. Rev}, 151(1):258--264, 1966.

\end{thebibliography}

\end{document}